%% file: main.tex
\documentclass[twocolumn,nolinenumbers,trackchanges]{aastex7}
\usepackage{graphicx} 
\usepackage{amsmath}
\usepackage[bottom]{footmisc}
\usepackage{hyperref}
\usepackage{booktabs}
\usepackage{placeins}

\begin{document}

\title{SED Fitting of the Globular Clusters in Abell 2744 at z=0.3}
\author[0009-0002-8949-4547]{Jinoo Kim}
\affiliation{Department of Physics \& Astronomy, McMaster University, 1280 Main St W, Hamilton, ON L8S 1T7, Canada}
\email[show]{kim697@mcmaster.ca}

\author[0000-0001-8762-5772]{William E. Harris}
\affiliation{Department of Physics \& Astronomy, McMaster University, 1280 Main St W, Hamilton, ON L8S 1T7, Canada}
\email[show]{harrisw@mcmaster.ca}

\begin{abstract}
In this study, we present a spectral energy distribution (SED) fitting analysis of globular cluster (GC) candidates in Abell 2744, a massive galaxy cluster at redshift $z = 0.308$ with a lookback time of 3.5 Gyr, using deep JWST/NIRCam photometry. By leveraging observations in eight broadband filters (F070W to F444W), we derive constraints on GC masses, ages, and metallicities of 69 GC candidates, refining previous studies focused on color-magnitude analysis.

The inferred GC masses predominantly fall near $10^7 M_\odot$, consistent with the high-mass end of typical GCs seen in the Local Group and overlapping with UCDs (Ultra-Compact Dwarfs). Although the best-fit age estimates are unavoidably uncertain, we very tentatively identify two prominent concentrations: one at intermediate ages (4-6.5 Gyr) near [M/H]=-0.25; and a second at older ages 8 - 10 Gyr  with both metal-poor ([M/H]=-0.66) and metal-rich ([M/H]=+0.15) components.

As expected for broadband SED fitting, we find mass-age and age-metallicity degeneracies in our result. Monte Carlo error propagation and tests with different SSP model assumptions show that the population-level trends in age and metallicity are preserved, although individual age estimates must be interpreted with caution. The resulting mass-to-light ratios show wavelength dependence with a mild decline toward longer wavelengths and increase with age, while showing minimal sensitivity to metallicity.

These results demonstrate the power and current limitations of multi-band JWST SED fitting that highlight the potential of future JWST data for detailed characterization of GCs at lookback times of several Gigayears.

\end{abstract}

\section{Introduction} \label{sec:Intro}

Globular clusters (GCs) are crucial to understanding the formation and evolution of galaxies, particularly in the early universe. Studying GCs at intermediate redshift provides unique insights into the processes governing galaxy assembly and star formation at look-back times of several billions of years. The importance of this research is amplified by the capabilities of the James Webb Space Telescope (JWST), whose unparalleled sensitivity and resolution in the near-infrared allow for detailed photometric analysis of GCs in such distant environments.

The Abell 2744 galaxy cluster, also known as Pandora's Cluster, is one of the most massive and complex clusters known, characterized by a dynamic merging history \citep[e.g.][]{merger1,merger2}, a rich diversity of galaxies \citep[e.g.][]{galaxy_pop}, and its substructures \citep[e.g.][]{substructure2,merger/substructure1}. It has been the target of extensive JWST observations compiled by the UNCOVER team, which integrates imaging data from multiple programs (\citealt{UNCOVER}).
The UNCOVER survey provides exceptionally deep NIRCam imaging of $\sim$45$\ \text{arcmin}^2$ around Abell 2744, reaching depths of $m_{AB}\sim$30, making it among the deepest JWST datasets available. The GCs within Abell 2744 are of particular interest because they represent an earlier stage of GC evolution at a redshift of z = 0.308, providing an opportunity to investigate characteristics of GCs 3.5 Gyr ago. This amazing dataset provides an invaluable window into the characteristics of extragalactic GCs.

Despite the advent of deep imaging, most extragalactic GC studies have historically relied on only two broadband filters to infer colors, metallicities, and luminosity functions \citep[e.g.][]{two_bands1,two_bands2,Harris_two_filters}. This two-band approach remains popular because it maximizes survey efficiency and sample size. As a result, there has been relatively less focus on full spectral energy distribution (SED) fitting across multiple filters \citep[e.g.][]{Previous_SED_fit2,Previous_SED_fit1,hartman+2025}. Moreover, previous work done on SED fitting of extragalactic GCs has focused on nearby galaxies \citep[e.g.][]{M31_SED, M81_SED, M87_SED, Virgo_SED}, with even less attention on GC populations at higher redshifts.

Previous studies of Abell 2744 \citep{H&RCabell2744_1,H&RCabell2744_2} have utilized JWST NIRCam Short Wavelength Channel (SWC) photometry to explore the luminosity functions, color distributions, and spatial characteristics of over 10,000 GCs in Abell 2744. By employing a broader set of NIRCam filters (F070W, F090W, F115W, F150W, F200W, F277W, F356W, and F444W), our work extends these studies, offering a more comprehensive analysis of the GCs' SEDs. This approach allows for more accurate determinations of their ages and metallicities, which are essential for constraining their star formation histories and chemical enrichment processes. Because this study focuses on the brightest unresolved point sources, their inferred masses overlap with the regime commonly associated with ultra-compact dwarfs (UCDs) (see Table \ref{tab:gc_photometry} for inferred masses). We therefore regard the present sample as the massive end of the GC-candidate population, but likely including objects with UCD-like properties. Throughout this paper, we use the term GC candidates to refer to the selection method, while recognizing that some objects may be better described as massive GCs or UCD-like compact stellar systems. For more detailed discussion on UCD populations in Abell 2744, see Section 7 of \cite{H&RCabell2744_1}.

The structure of this study is presented as follows: In Section \ref{sec:Data}, the observed data of Abell 2744 and the simple stellar population (SSP) model spectra used in this research are explained. In Section \ref{sec:SED}, we explain how to convert the observed data for the fitting purpose and the method of SED fitting. Section \ref{sec:results} shows the results of the fitting.
Section \ref{sec:discussion} discusses the implications and validity of these results, while Section \ref{sec:conclusion} summarizes our key conclusions and highlights prospects for future JWST‑based GC studies.

We adopt cosmological parameters $H_0 = 67.8\text{km s}^{-1}$\footnote{Having a higher Hubble constant such as $H_0=74\text{km s}^{-1}$ would reduce the distance by 8\% and the luminosities and masses by 18\%} and $\Omega_{\Lambda} = 0.692$, which for $z = 0.308$ gives a luminosity distance $d_L = 1630$ Mpc for Abell 2744, and an angular size distance of $d_A = d_L/(1 + z)^2 = 952$ Mpc. The foreground extinction (from the NED database) is $A_V = 0.036$, which has a negligible effect for the NIR bandpasses used in this study.

\section{Data} \label{sec:Data}
\subsection{Abell 2744} \label{subsec:abell2744_data}
In this study, we use the mosaic images from Data Releases DR2 and DR3 of the UNCOVER survey \citep{UNCOVER}. The dataset includes photometry from a broad range of NIRCam filters covering both the SWC and Long-Wavelength Channel (LWC). Specifically, the NIRCam filters used in this work span from F070W to F444W. This comprehensive wavelength range is essential for accurately sampling the SEDs of the GCs and for performing detailed age and metallicity fitting using  SSP models (see below).

For photometric measurements, we applied the point spread function (PSF) photometry techniques with the DAOPHOT package \citep{Daophot} in its \emph{iraf} implementation. The major cleaning steps using DAOPHOT parameters to isolate the point sources in the field (which are dominated by GC and UCD candidates) took place in the photometry from the SWC filters. This was described completely in \citet{H&RCabell2744_1} and \citet{H&RCabell2744_2}. The candidate list of GCs from that study was initially identified from F115W, F150W, and F200W images due to their similar limiting magnitudes and resolution. These three SEC filters are the deepest images of all the filters in both SWC and LWC, and also have the advantage of higher resolution than the LWC bands, which allowed better selection of point sources versus contaminants. We then adopted this list for measurement on all the 8 broadband filters in this study. PSFs were derived separately for each filter, then the aperture corrections were made in two steps, from the PSF fitted magnitudes out to 10 px aperture, and then from there out to large radius with the published Encircled Energy curves for each filter. The globular cluster luminosity function (GCLF) was discussed in \citet{H&RCabell2744_2}, along with estimates of the total GC population. The estimated GCLF turnover point is at $m_{AB} = 32$, which is far fainter than the limits of either the SWC or LWC images. The limiting magnitudes vary between filters, with the LWC bands being shallower than the SWC bands, but the brightest GCs are well measured across all bands. The detailed process of constructing the catalog in the LWC will be provided in a follow-up paper.

\begin{figure}[!ht]
    \centering
    \includegraphics[width=0.47\textwidth]{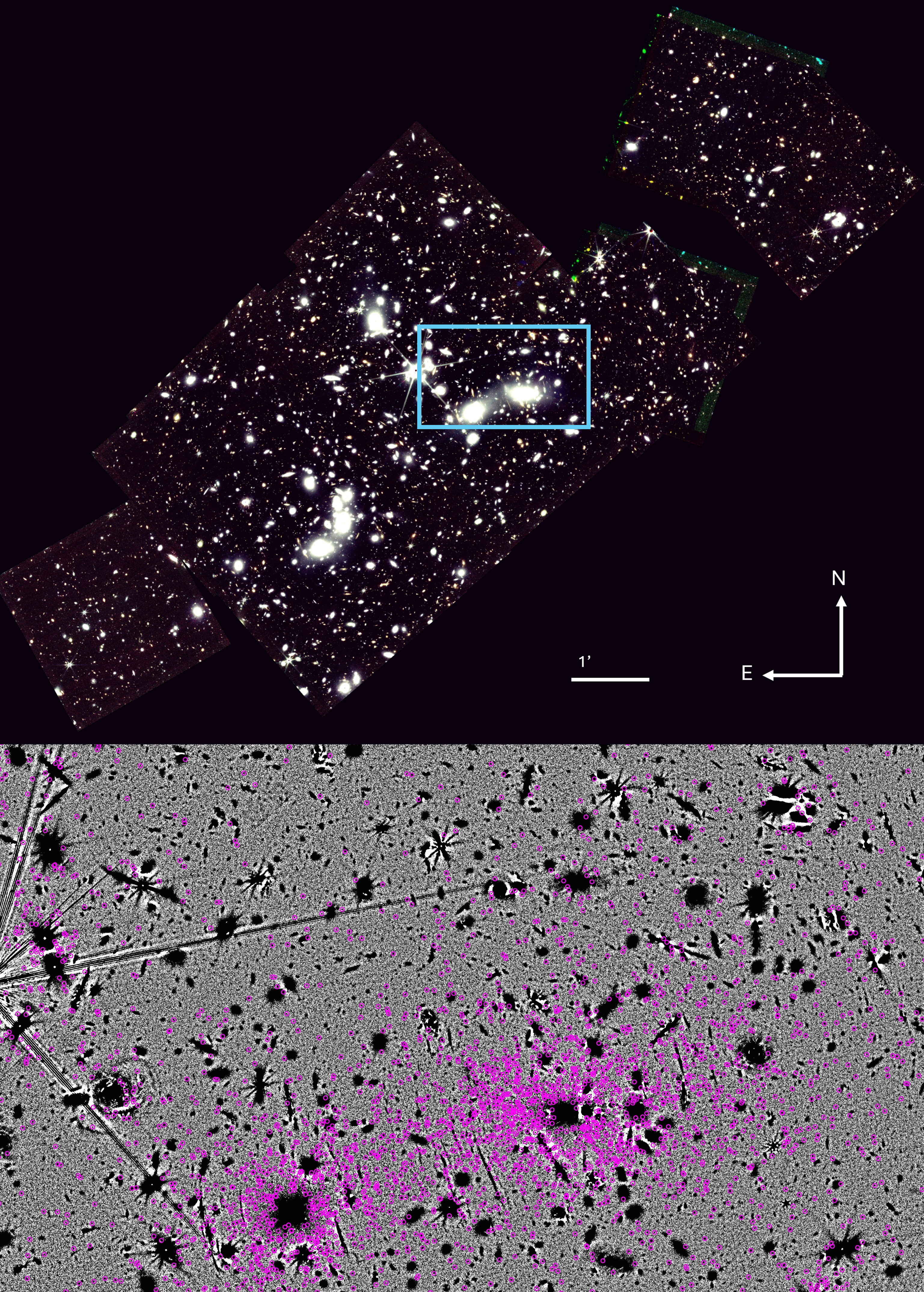} \caption{\textit{Top}: RGB image of Abell 2744. Up is north and left is east. \textcolor{red}{Red}: F444W, \textcolor{green}{Green}: F356W, \textcolor{blue}{Blue}: F277W. \textit{Bottom}: A zoomed-in image of the blue rectangle of the top figure with GCs marked in magenta circles. The image has been median filtered to remove large-scale gradients in background light.}
    \label{fig:abell2744}
\end{figure}

The image of GC candidates in Abell 2744 is shown in Figure \ref{fig:abell2744}.

\subsection{Model Spectra}

\begin{table*}[!htb]
\centering
\caption{Adopted fiducial SSP model grid used for the $\chi^2$ minimization SED fitting.}
\label{tb:age_mh}
\begin{tabular}{ll}
\hline
Parameter & Grid values \\
\hline
Age (Gyr) & $0.5$--$10.0$ for baseline grid and $0.5$--$13.0$ for extended grid in steps of $0.5$ Gyr\\
$\mathrm{[M/H]}$ & $-2.27,\,-1.79,\,-1.49,\,-1.26,\,-0.96,\,-0.66,\,-0.35,\,-0.25,\,+0.06,\,+0.15,\,+0.26$ \\
\hline
\end{tabular}
\end{table*}

We utilize synthetic spectra generated from the E-MILES stellar population models, which provide SEDs for SSPs 
(Simple Stellar Populations)
over a wide range of ages and metallicities. These models are particularly well-suited for our purposes, as they offer high-resolution spectra that span from 1680 \AA \ to 50000 \AA \citep{emiles_code} and therefore cover all the normally used photometric bands that we need. 

To generate the synthetic spectra, we initially used BaSTI isochrones \citep{BaSTI} to model the stellar evolutionary stages and the Chabrier Initial Mass Function \citep{Ch_IMF} to account for the initial stellar mass distribution. Together, these models were incorporated into the E-MILES framework, which allowed us to generate SEDs over all the combinations of age and metallicity listed in Table \ref{tb:age_mh}.  Note that GC ages larger than 10 Gyr would not be permitted for the lookback time of Abell 2744, while the metallicity values cover the range of most of the observed GCs in nearby galaxies \citep[e.g.][]{GC_age1,GC_MH1,GC_MH_age1,GC_M31}. The major results shown in Section~\ref{subsec:main_result} below are based on the age range up to 10 Gyr. However, the model grid in Section~\ref{subsec:larger_grid} includes ages up to 13 Gyr strictly as a numerical test to investigate how the choice of the model grid range affects the fitting results. 

In addition to the fiducial E-MILES grid constructed with BaSTI isochrones and Chabrier IMF, we also generated several comparison grids to test the sensitivity of the fitted parameters to SSP model assumptions. Specifically, we repeated the fitting with BaSTI isochrones and bimodal IMFs with slopes 0.80 and 2.30, as well as with a Padova isochrone set \citep{Padova00} with Chabrier IMF. These alternative model sets were processed through the same throughput-weighting and fitting process as the fiducial grid. The BaSTI-Chabrier models are retained as the primary results of this work, while the additional model variants are used as robustness tests to evaluate whether or not the inferred parameters are sensitive to the adopted SSP prescription.

These SSP models account for stellar evolution within a single-age, single-metallicity  framework, including changing luminosity of stars as they evolve off the main sequence. However, they do not account for dynamical evolution of a bound star cluster. Processes such as tidal stripping, evaporation, and mass segregation can preferentially remove low-mass stars in real GCs, changing the present-day mass function and reducing the mass-to-light ratio relative to a purely stellar-evolution model. For the Abell 2744 GC population, \cite{H&RCabell2744_1} used a first-order dynamical mass-loss prescription to estimate that evolution from z = 0.308 to z = 0 would remove $\sim15\%$ of the mass at the high-mass end that we observe here, with the exact amount depending on the cluster mass, orbit, and local gravitational potentials. Because these quantities are not constrained for individual candidates in this study, we do not attempt to apply an object-by-object dynamical correction. Therefore, the masses derived in this work should be interpreted as photometric stellar population masses tied to the adopted SSP model assumptions.

\begin{figure}[!htbp]
    \includegraphics[width=0.47\textwidth]{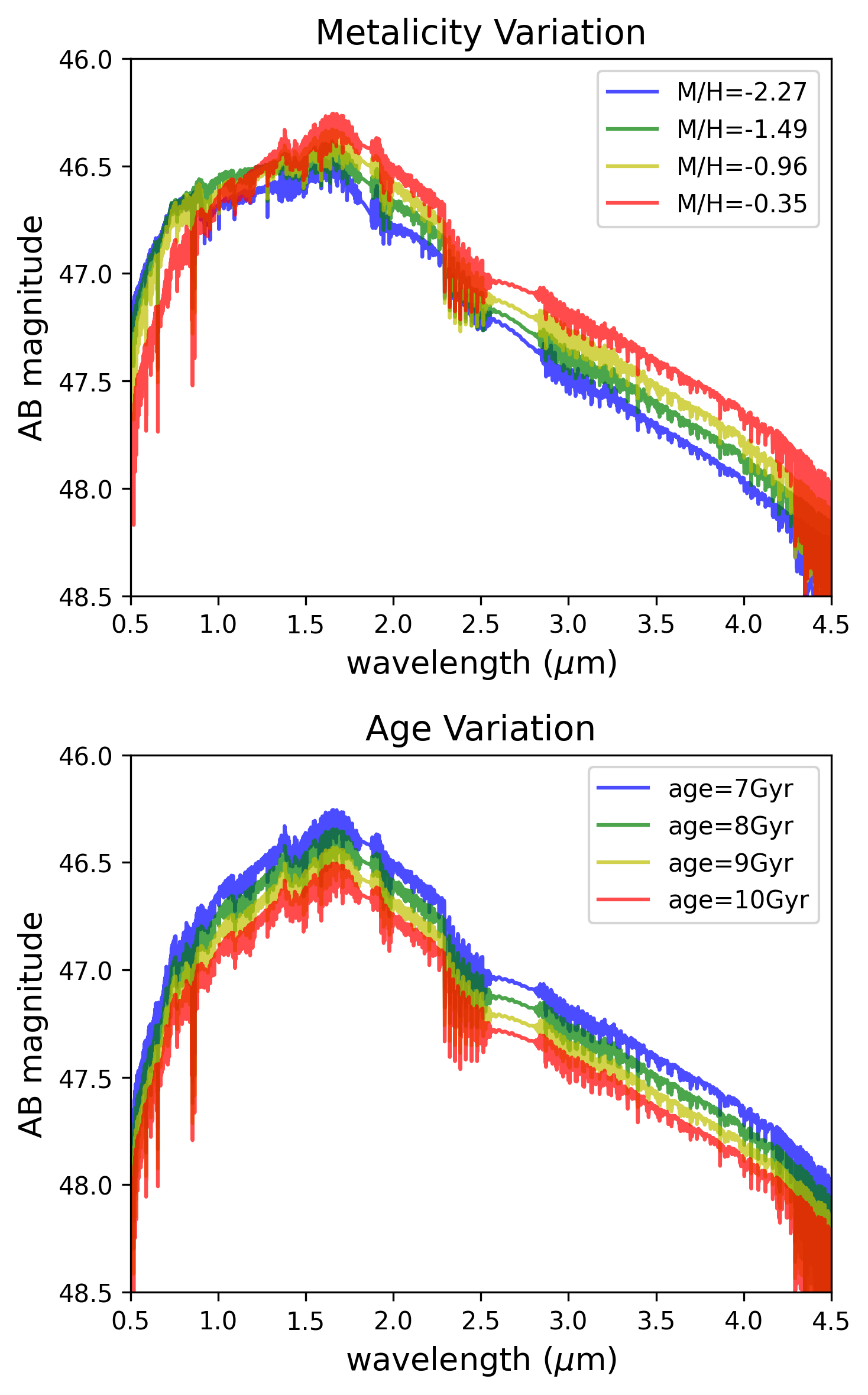}
    \caption{Model spectra (per solar mass) in the same source plane as Abell 2744. \textit{Top}: Variations on the fiducial model spectra when age is fixed as 7 Gyr and metallicities are varied. \textit{Bottom}: Variations on the fiducial model spectra when metallicity is fixed as [M/H] = -0.35 and ages are varied.}
    \label{fig:SED_variation}
\end{figure}

Examples of E-MILES model spectra, shown in Fig.~\ref{fig:SED_variation}, 
illustrate the SEDs under varying metallicities and ages, highlighting how these physical properties influence the observable characteristics of the spectra. The synthetic spectra are given in units of Solar luminosity per Angstrom per Solar mass, effectively corresponding to an \textit{AB magnitude per solar mass}, which differs from standard AB magnitudes by an offset of $2.5 \log_{10}(M_{GC})$, as detailed in Section \ref{subsec:conversion}. In the top panel of the figure, increasing metallicity (from [M/H] = -2.27 to -0.35) causes a redward shift in the spectra, characterized by enhanced flux at longer wavelengths and diminished flux at shorter wavelengths. Meanwhile, the bottom panel shows that variations in the age of the stellar population (7 - 10 Gyr) primarily affect the overall flux level with only minor changes in the shape of the spectra. Hence, we expect our SED fitting to be more sensitive to metallicity than to age.

\section{SED fitting} \label{sec:SED}
\urldef{\myurl}\url{https://jwst-docs.stsci.edu/jwst-near-infrared-camera/nircam-instrumentation/nircam-filters#NIRCamFilters-Filtertransmissions}
\subsection{Data Conversion} \label{subsec:conversion}
We convert the model spectra to the observational plane and fit directly to the measured imaging data as in other recent literature on SED fitting.

E-MILES gives spectra of SSPs in units of Solar luminosity per Angstrom per Solar mass. In order to compare this to the apparent magnitude of the samples in Abell 2744 we need to account for the effects of the distance and masses of the GCs. Then the observed flux $f_{\lambda}$ can be expressed as
\begin{equation} \label{eq:observed_flux}
f_\lambda = \frac{M_{\text{GC}} L_\odot}{4 \pi d^2} F_\lambda
\end{equation}
where $M_{GC}$ is the cluster mass in solar mass, 
and $d = 1630 \textrm{Mpc} = 5.03 \times 10^{27} \textrm{cm}$ is the luminosity distance to Abell 2744, and $F_\lambda$ is the E-MILES flux.




In the AB magnitude system, Eq.\ref{eq:observed_flux} can be converted to
\begin{multline} \label{eq:simplified_formula}
m_{\text{AB}} = 38.226 + 5 \log_{10}(d) - 5 \log_{10}(\lambda) \\ - 2.5 \log_{10}(F_\lambda) - 2.5 \log_{10}(M_{\text{GC}})
\end{multline}
where d is now in Mpc, $\lambda$ is in \AA, $F_{\lambda}$ is in Solar luminosity per \AA\  per Solar mass, and $M_{GC}$ is in Solar masses.

\subsection{K-correction}\label{sec:k_correction}
\begin{figure}[!ht]
    \centering
    \includegraphics[width=0.47\textwidth]{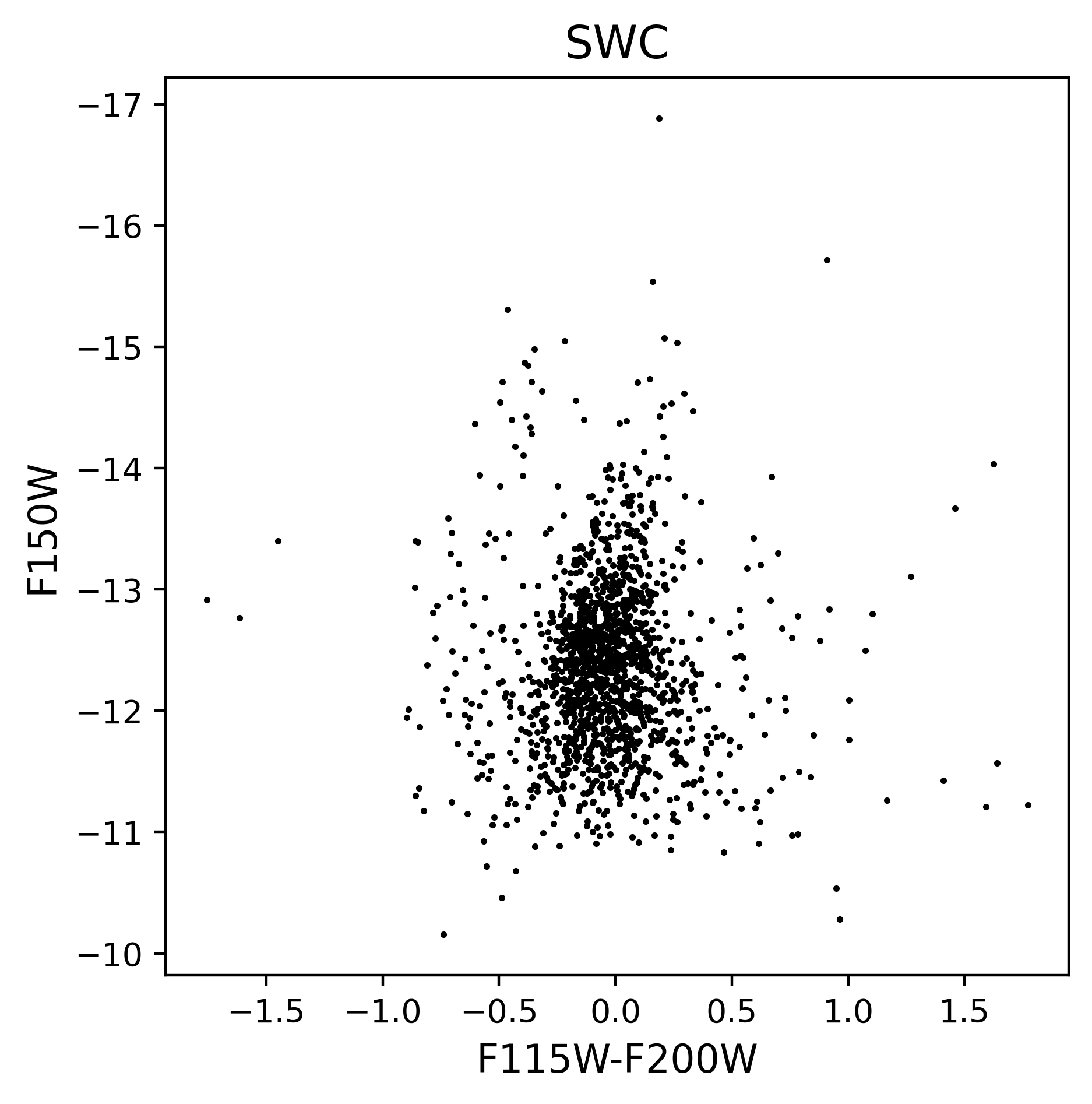}
    \includegraphics[width=0.47\textwidth]{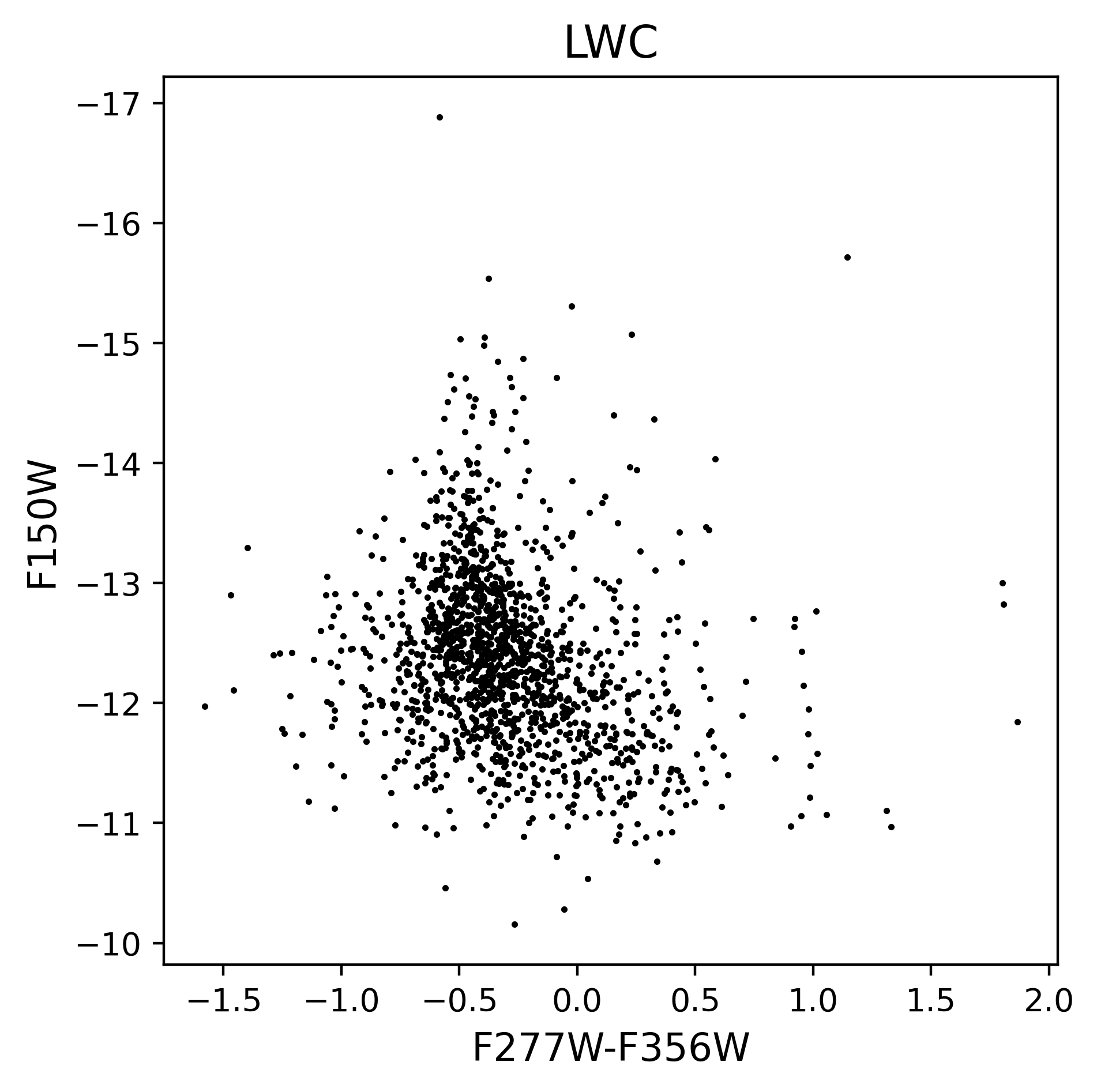} \caption{Color magnitude diagrams from absolute magnitudes of LWC and SWC accounting for distance and K-correction. There is a total of 1495 objects present in all filters.}
    \label{fig:CMD_all}
\end{figure}

Because Abell~2744 lies at z = 0.308, the observed NIRCam photometry samples bluer rest-frame wavelengths, and an additional bandpass-dependent correction is required whenever we convert apparent magnitudes into rest-frame absolute magnitudes. The SED fitting process itself in this study does not require K-correction values, as we are fitting the model spectra to apparent magnitudes, but the K-correction is needed for the analysis or quantities that depend on absolute magnitudes such as Fig.~\ref{fig:CMD_all} and mass-to-light ratios in Section~\ref{sec:ML_ratio} (Fig.~\ref{fig:ML_ratio}). K-correction accounts for how a source’s emitted spectrum is shifted and dimmed or brightened by redshift before it reaches us through a given filter. In practice, it answers how much we should adjust the observed apparent magnitude so that it represents the intrinsic (rest-frame) magnitude in the same bandpass. The exact values of the K-corrections of SSP models at different ages and metallicities at each filter are calculated by using the homochromatic K-correction of the \texttt{RESCUER} webtool \citep{k_correction}.

The absolute magnitude of an object can be calculated as:
\begin{equation} \label{eq:absolute_mag}
    \text{M}=m-5\text{log}_{10}\left(\frac{d}{10\text{pc}}\right)-K
\end{equation}
where M is absolute magnitude in a given filter, $m$ is apparent magnitude, $d$ is distance to the object in pc, and $K$ is K-correction of the object. 
The exact K-correction depends on the adopted SSP age and metallicity, which are themselves affected by the mass-age and age-metallicity degeneracies inherent to the fitting (the degeneracies are further explained in Section~\ref{subsec:mass_age_degeneracy} and \ref{subsec:age_mh_degeneracy}). However, across the range of neighboring models permitted by these degeneracies, the K-correction varies only modestly, only $\leq 0.05$. Thus, while the application of K-corrections is necessary, the small model-to-model differences in K-correction do not significantly alter the the conclusions of this work. We therefore regard the adopted K-correction values as robust and safely applicable for the purposes of this study.

\subsection{Effect of Throughput}
To accurately compare model spectra with observed photometry, it is essential to account for the effects of the filter transmission across the full bandpass. Although the pivot wavelength provides a representative comparison point, photometric observations inherently reflect integrated flux across the entire filter range. Therefore, both a characteristic wavelength and a band-averaged model magnitude must be computed.

The effective wavelength serves as a single representative wavelength for each filter that reflects the combined influence of the filter transmission function and the SED of the source. Following the definition in \citet{Isophotal}, it is computed as:

\begin{equation} \label{eq:effective_wavelength}
    \lambda_{\text{eff}} = \frac{\int \lambda S(\lambda) T(\lambda) \, d\lambda}{\int S(\lambda) T(\lambda) \, d\lambda} \approx \frac{\sum \lambda_i S(\lambda_i) T(\lambda_i) \, \Delta\lambda}{\sum S(\lambda_i) T(\lambda_i) \, \Delta\lambda},
\end{equation}
where $S(\lambda)$ represents the source flux density, and $T(\lambda)$ denotes the total system throughput, including the filter transmission function and instrumental response. The product $S(\lambda)T(\lambda)$ effectively acts as a weighting function, determining the probability of detecting an emission from the source at wavelength $\lambda$. Consequently, $\lambda_{\text{eff}}$ corresponds to the mean wavelength weighted by the energy distribution of the source across the filter.

The computed effective wavelengths for the NIRCam filters used in this study are listed in Table~\ref{tb:effective_wavelength}.  Here we have used a sample SED model spectrum with the age of 8 Gyr and metallicity of [M/H] = -0.96. We note that for all but the reddest filter F444W, $\lambda_{eff}$ is closely similar to the pivot wavelength $\lambda_{pivot}$ listed in the NIRCam webpages.

\begin{table}[]
\centering
\caption{Effective wavelengths for the NIRCam filters used in this study. Pivot wavelengths are taken from \url{https://jwst-docs.stsci.edu}, and effective wavelengths are derived using Eq.~\ref{eq:effective_wavelength}. All values are in $\mu$m.}
\label{tb:effective_wavelength}
\begin{tabular}{lll}
\hline
Filter & $\lambda_{\text{pivot}}$ ($\mu$m) & $\lambda_{\text{eff}}$ ($\mu$m) \\ \hline
F070W  & 0.704 & 0.706 \\
F090W  & 0.903 & 0.903 \\
F115W  & 1.154 & 1.147 \\
F150W  & 1.501 & 1.494 \\
F200W  & 1.988 & 1.977 \\
F277W  & 2.776 & 2.726 \\
F356W  & 3.566 & 3.524 \\
F444W  & 4.401 & 4.334 \\ \hline
\end{tabular}
\end{table}

Since the observed magnitudes are observations over entire bandpasses rather than being measured at a single wavelength, the model-predicted magnitudes must also be computed as band-averaged values, accounting for the full filter response.

To facilitate direct comparisons, we compute the band-averaged magnitude per solar mass from the model spectra. This quantity, before accounting for the GC mass, can be expressed as:
\begin{equation}
\begin{aligned} \label{eq:mag_avg_per_sol}
    m_{j, \text{avg}, \odot} = &38.226+5\text{log}_{10}(d)-5\text{log}_{10}(\lambda_{eff,j})\\&-2.5\text{log}_{10}\left(\frac{\sum(F_{\lambda_i}\cdot T({\lambda_i}))}{\sum T({\lambda_i})}\right)
\end{aligned}
\end{equation}
where $j$ indicates a broadband filter $j$, d is distance Mpc, $\lambda_{eff,j}$ is $\lambda_{eff}$ of the filter $j$, $F_{\lambda_i}$ is E-MILES flux at wavelength $\lambda_i$ and $T(\lambda_i)$ is the filter transmission function at wavelength $\lambda_i$. This averaged magnitude, incorporating the filter transmission function, is subsequently calculated for each filter for all model spectra in the SED fitting process.


\subsection{Fitting Method} \label{subsec:fitting}
To derive the best-fitting parameters for the GC candidates, we employ a standard minimization of the $\chi^2$ statistic, defined as:

\begin{equation} 
    \chi^2 = \sum_{j=1}^{N} \frac{(O_j - E_j)^2}{\sigma_j^2},
\end{equation}

where $O_j$ is the observed magnitude, $E_j$ is the estimated magnitude from the model spectra, $\sigma_j$ is the measurement uncertainty, and $N=8$ is the number of NIRCam filters used. Following Eq.\ref{eq:simplified_formula} and Eq.\ref{eq:mag_avg_per_sol}, the estimated magnitude $E_j$ accounting for the mass of the GC is calculated as:

\begin{equation} \label{eq:predicted_mag}
    E_j = m_{j,\text{avg},\odot} - 2.5 \log_{10}(M_{\text{GC}}),
\end{equation}

where $m_{j,\text{avg},\odot}$ is the band-averaged model magnitude per unit solar mass from Eq.\ref{eq:mag_avg_per_sol}, and $M_{\text{GC}}$ is the GC mass. The best fit is found by minimizing $\chi^2$ across all models.

The fitting procedure evaluates a grid of 220 models (286 for the larger grid test in Section~\ref{subsec:larger_grid}) for each candidate. This means for each candidate, there are 220 fitting procedures finding the masses that yield minimum $\chi^2$ for each model. Then the model with the minimum $\chi^2$ across all models is selected as the best fit, giving the best-fit age, metallicity, and mass of the GC candidate.

\subsection{Photometric Uncertainty Propagation} \label{sec:error_propagation}
The best-fit parameters in this study are first obtained from the grid-based $\chi^2$ minimization described before. To estimate the effect of photometric uncertainty on inferred age, metallicity, and stellar mass, we perform a Monte Carlo (MC) error propagation analysis. For each GC candidate, we generate 500 realizations of the observed eight band photometry by perturbing each magnitude according to a Gaussian distribution centered on the measured magnitude with a standard deviation equal to its DAOPHOT uncertainty. Each realization is then refit using the same fiducial SSP model grid. For each object, we use the 16th, 50th, and 84th percentiles of the resulting best-fit parameter distributions as the formal MC uncertainty range.

\subsection{Data Selection} \label{sec:data_selection}
\begin{figure}
    \centering
    \includegraphics[width=0.97\linewidth]{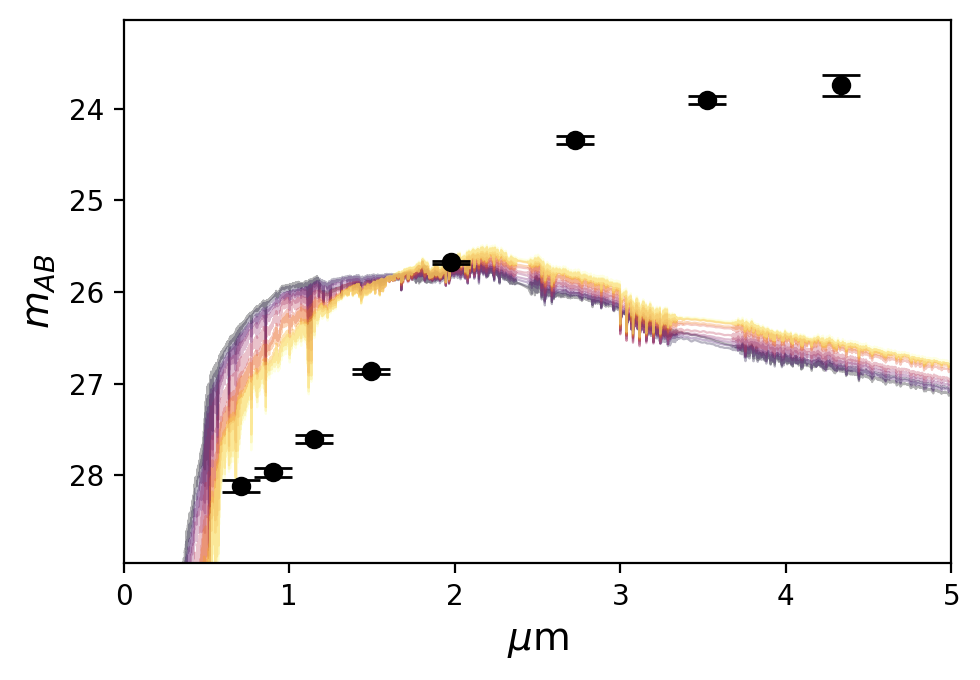}
    \caption{An example of GC candidates with exceptionally poor fit. The black dots are observed magnitudes and the solid lines in different colors represent SED of models with different metallicities. Clearly, the shape of the observed magnitudes does not resemble any of the model spectra.}
    \label{fig:bad_fit}
\end{figure}

To ensure the robustness of the SED fitting analysis, stringent criteria were applied to discard poor data. As mentioned in Section \ref{subsec:abell2744_data}, roughly $\sim$10,000 GC candidates were initially identified from the SWC images. Then only the GC candidates that appear in all eight NIRCam filters and have brightness $m_{AB, \text{F356W}} \leq 28.2$, which are well-above the estimate of the completeness levels of the LWC images from what magnitude the CMD drops off, are included for further analysis. Candidates with obviously poor fits to the model spectra, such as Fig.~\ref{fig:bad_fit}, are excluded.
The shape of the observed magnitudes of the poor fits shows clear deviations from all of the model spectra mostly with a mismatch to the turnover brightness (magnitudes) around 2 $\mu$m. Potential causes for poor fits include misclassification as background galaxies or foreground stars, contamination from background light, unresolved blending with nearby sources, simple measurement scatter, or even undetected variability.

Further refinement was performed by excluding candidates with exceptionally high $\chi^2$ values. While most candidates exhibited similar minimum $\chi^2$ values, a few outliers displayed values orders of magnitude larger. To identify and exclude these outliers, the mean absolute deviation (MAD) of the $\chi^2$ distribution was computed, and any candidate exceeding the mean $+\ 5\ \times$ MAD threshold $\approx 135$ was removed.

\begin{figure}[!t]
    \includegraphics[width=0.47\textwidth]{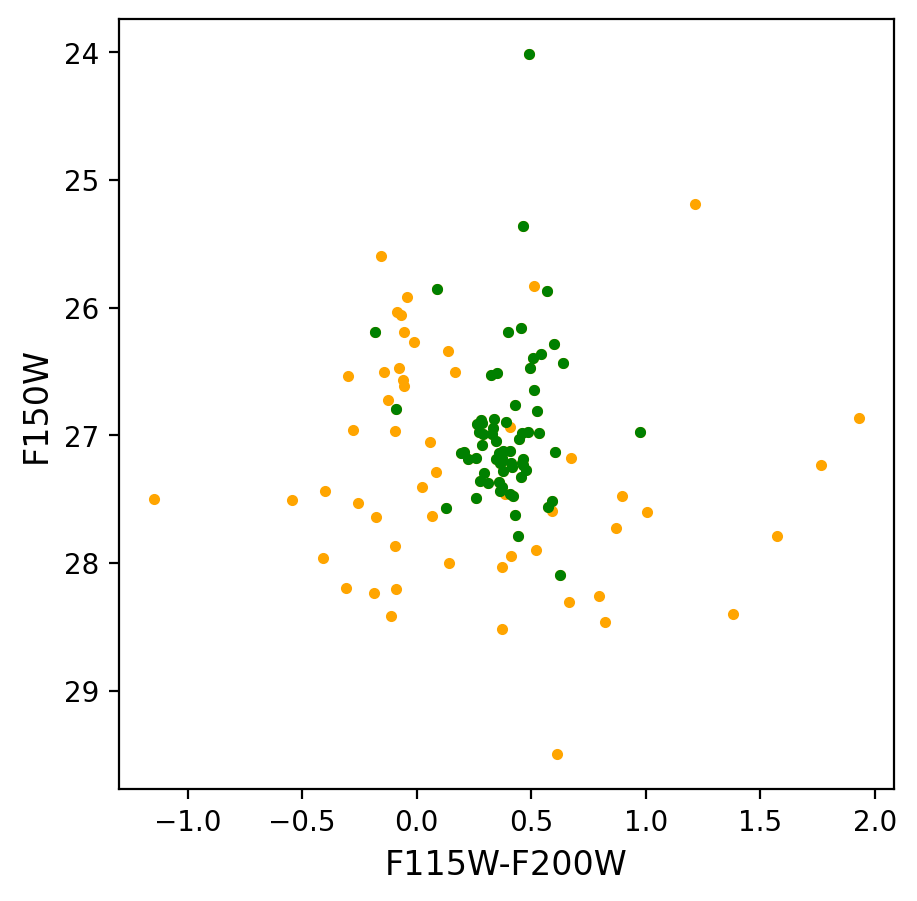}
    \caption{Color magnitude diagram of GC candidates that are reasonably bright ($m_{AB,F356W} \leq 28.2$) and present in all 8 filters. Green dots are the candidates with reasonable SED shapes, whereas orange dots are with unrealistic SED shapes. Magnitudes here are apparent magnitudes.}
    \label{fig:CMD_poor_fits}
\end{figure}

Excluded and included objects in the final sample are shown in Fig.~\ref{fig:CMD_poor_fits}, showing the CMD of GC candidates. The included GCs form a narrower sequence, while poorly fitting candidates (orange points) are primarily outliers.
After applying these criteria, a refined dataset of 69 GC candidates remained.

\begin{figure*}[!ht]
    \includegraphics[width=0.97\textwidth]{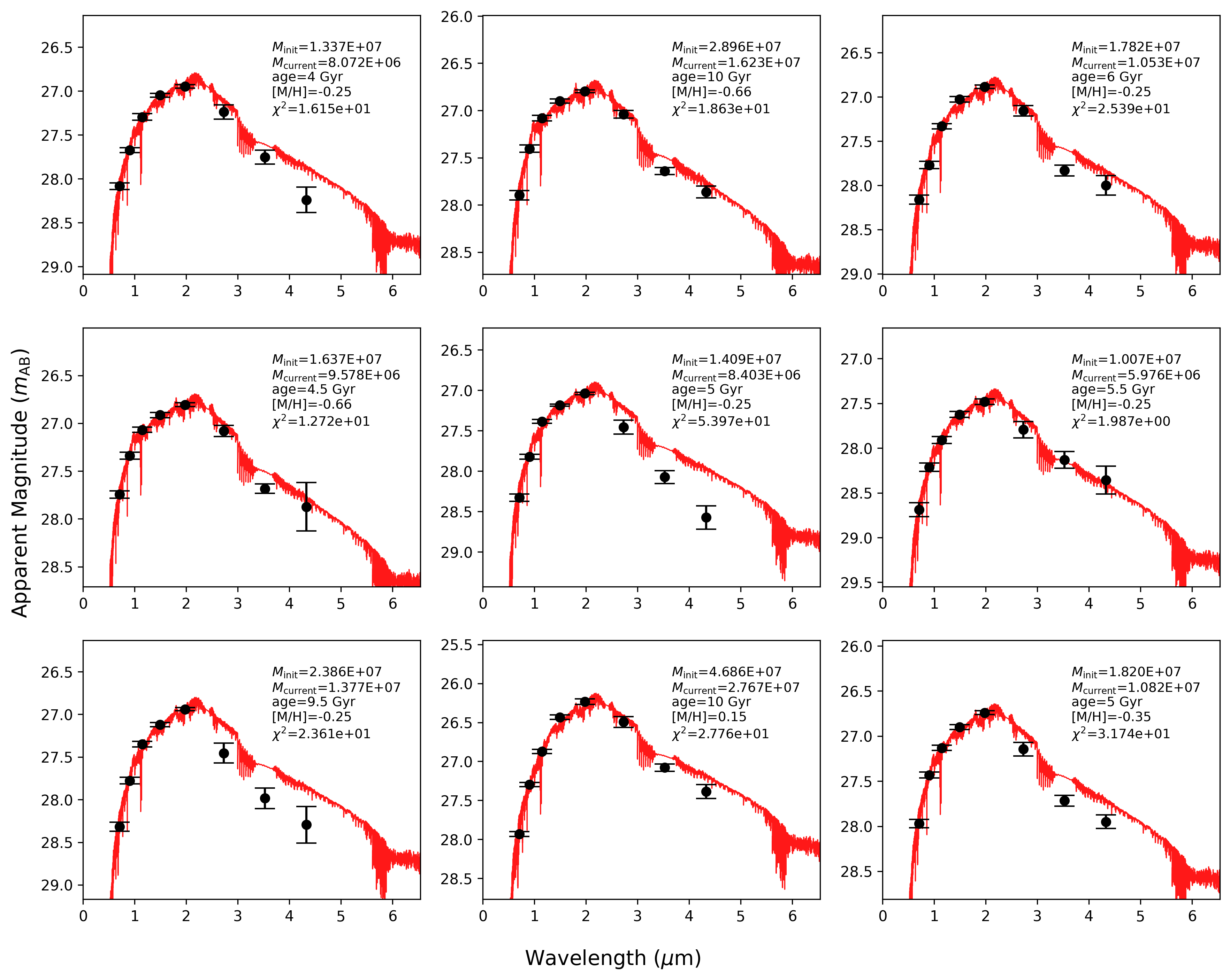}
    \caption{Examples of SED fitting in this study. The red lines are model spectra corresponding to the mass, age, and metallicity described in the figure. The black dots are measured magnitudes of each GC at different filters with error bars. Mass is in units of solar mass and $\chi^2$ value is also presented to show the goodness of the fit. $M_{\text{init}}$ is best-fit mass at the formation of the cluster and $M_{\text{current}}$ is currently observed best-fit mass (i.e. mass at z = 0.308).}
    \label{fig:Good_fits_mosaic}
\end{figure*}

\section{Results} \label{sec:results}
\subsection{Mass, Age, and Metallicity} \label{subsec:main_result}

\begin{figure*}[!ht]
    \includegraphics[width=0.97\textwidth]{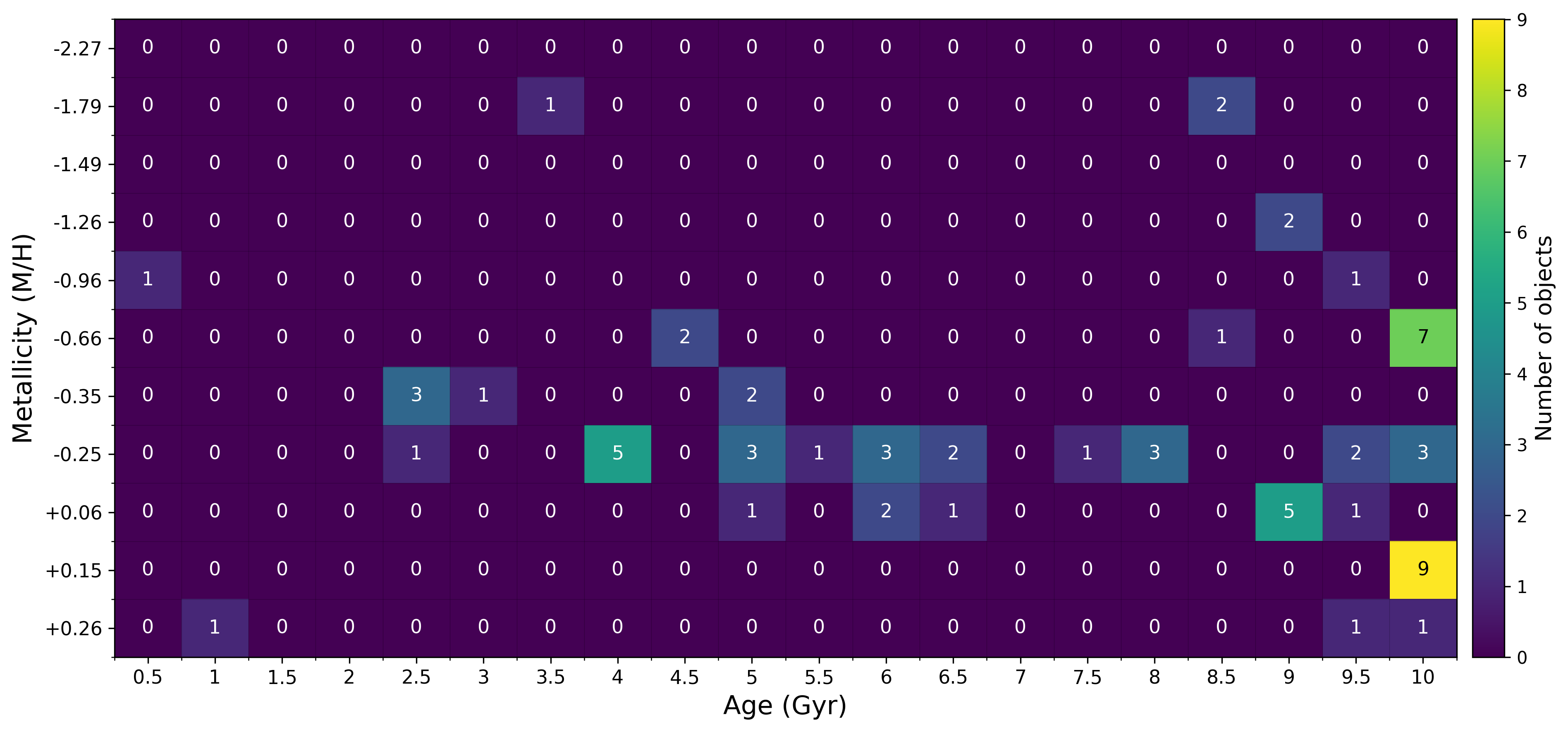}
    \caption{Frequency distribution (heatmap) of best-fit models by age and metallicity.}
    \label{fig:heatmap_upto10}
\end{figure*}

Fig.~\ref{fig:Good_fits_mosaic} shows a representative sample of SED fittings performed in this study. The figure illustrates that the observed eight JWST/NIRCam broad bands are generally well-reproduced by single SSP models across our wavelength baseline, yielding typical masses of order $10^7 M_{\odot}$ (mass at z = 0.308, denoted as $M_{\text{current}}$ in the figure). This inferred mass is consistent with the upper mass limit of typical GCs in bright cluster galaxies, or with typical masses of UCDs \citep{mass_limit}. This outcome aligns with expectations, as only the brightest GCs are detectable at this redshift due to observational limits. The exclusion of fainter sources, as detailed in Section \ref{sec:data_selection}, also ensured that brighter (hence more massive) sources were picked up to focus on higher signal-to-noise detections.

The distribution of best-fit solutions in age and metallicity is summarized in Fig.~\ref{fig:heatmap_upto10}. The resulting heatmap is clumpy rather than uniform. A substantial number of the GC candidates are located at the upper edge of the adopted age range (10 Gyr) with a pronounced metallicity bimodality at that age (9 objects at $\mathrm{[M/H]}=+0.15$ and 7 objects at $\mathrm{[M/H]}=-0.66$). Although clustering at the boundary of the model grid could indicate a fitting artifact rather than a physical feature, our tests with an expanded model grid (Section~\ref{subsec:larger_grid}) support the conclusion that this clustering is genuine. Additional concentrations occur at intermediate ages and moderate metallicity, across 4-6.5 Gyr at $\mathrm{[M/H]}=-0.25$. Only a small fraction of candidates occupy the most metal-poor bins, which, when present, preferentially appear at older ages. This is expected as metal-poor GCs tend to be older and thus fainter, and have maximum masses lower than for the metal-rich objects \citep{younger_metal_rich,H&RCabell2744_2}.

\begin{figure}[!htpb]
    \centering
    \includegraphics[width=0.95\linewidth]{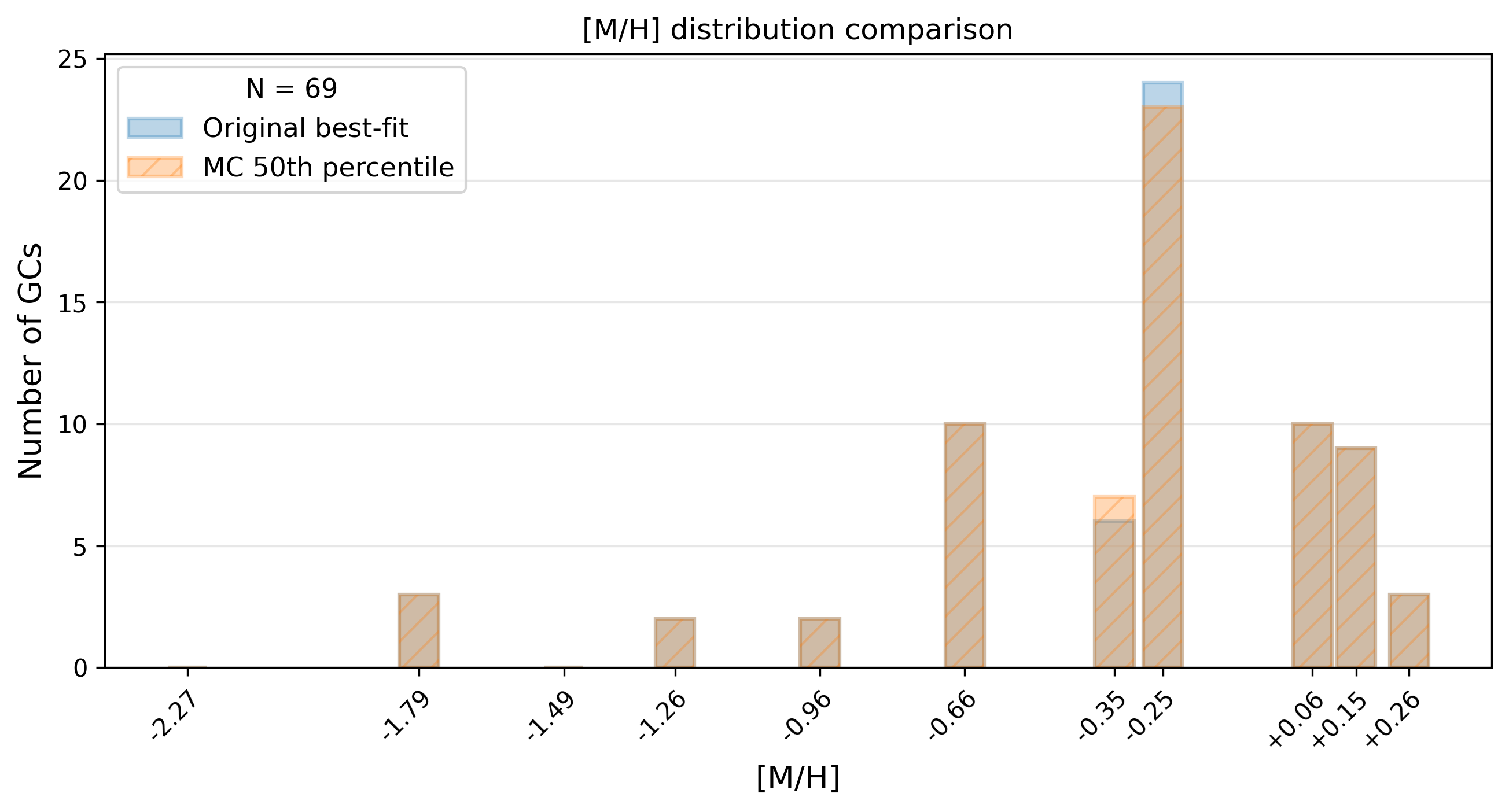}
    \includegraphics[width=0.95\linewidth]{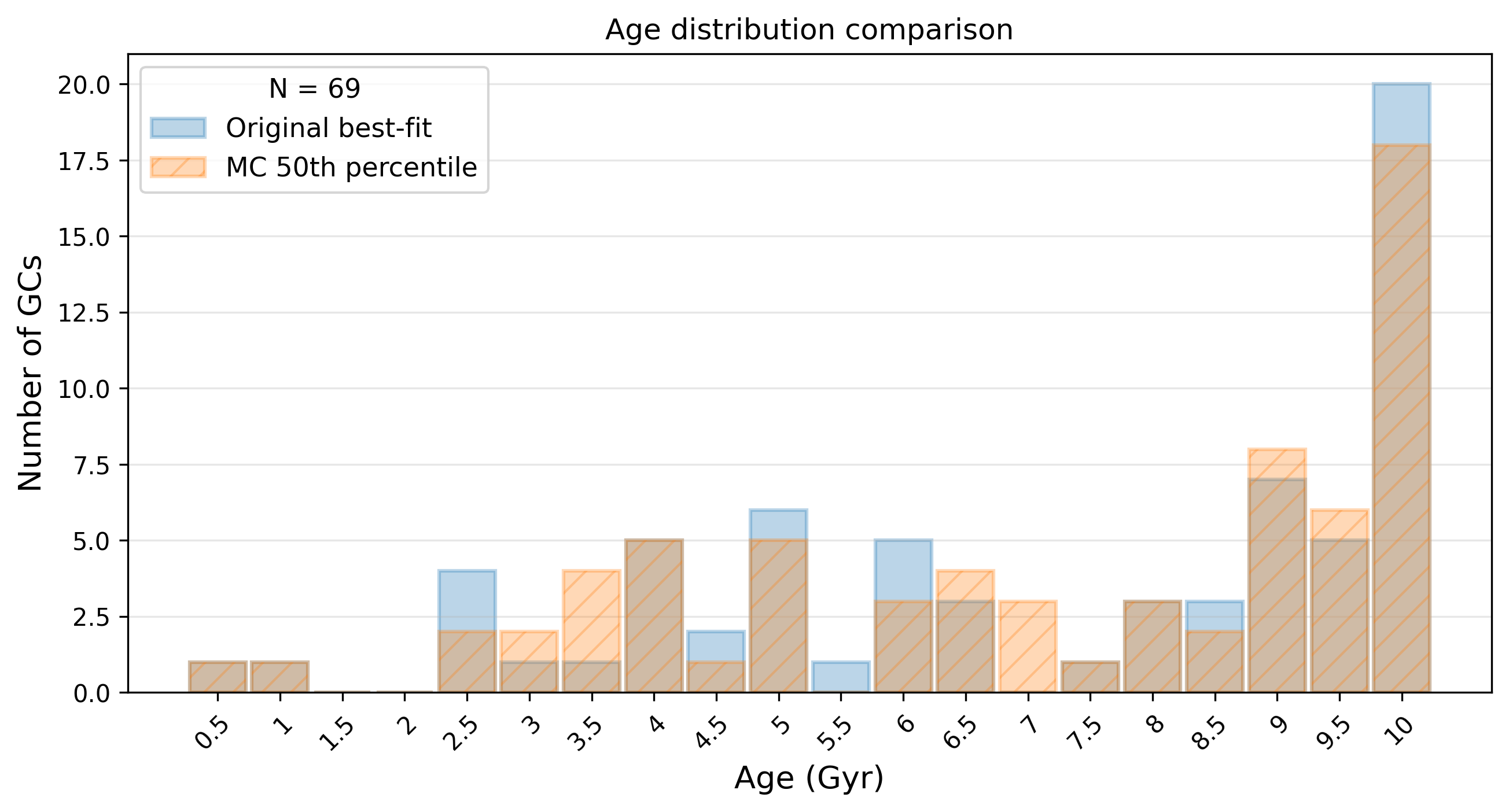}
    \caption{Comparison between the original best-fit parameter distributions and the results from MC photometric error propagation. Blue histograms show the best-fit values obtained from the original photometry, while orange hatched histograms show the 50th-percentiles from 500 MC realizations per object.}
    \label{fig:MC_propagation}
\end{figure}

The stability of these distributions is tested with MC photometric error propagation. In each realization, the eight observed magnitudes were perturbed within their measured uncertainties and the full SED fitting procedure was repeated. The resulting distributions are shown in Fig.~\ref{fig:MC_propagation}, where the original best-fit values are compared with the 50th-percentile values from the MC realizations. The metallicity distribution is largely preserved, with changes occurring between adjacent metallicity bins. The age distribution shows broader redistribution. This behavior is expected because broadband SED fitting constrains age less sharply than metallicity once the mass normalization is allowed to vary (see Section~\ref{subsec:mass_age_degeneracy} and \ref{subsec:age_mh_degeneracy}). The MC results therefore support the main features of the best-fit distribution, while emphasizing that individual ages should be interpreted with considerable caution.

The uncertainties in fitted parameters are shown in Table~\ref{tab:gc_photometry}. These derived uncertainties should be interpreted as formal photometric uncertainties propagated through the adopted SED-fitting pipeline. They quantify how the best-fit age, metallicity, and stellar mass change when the observed magnitudes are perturbed within their measured DAOPHOT errors, while keeping the SSP model grid, IMF and isochrones fixed. Therefore, these percentile ranges do not represent the full systematic uncertainty. In particular, they do not include uncertainties associated with the finite spacing of model grid, source contamination, and possible limitations of stellar population models. The reported MC uncertainties should therefore be regarded as lower limits on the total uncertainties, but they provide a useful measure of the sensitivity of the fitted parameters to the observed photometric errors.

The reduced $\chi^2$ values listed in Table~\ref{tab:gc_photometry} are computed as $\chi^2_\nu=\chi^2/\nu$, where $\nu=8-1=7$ corresponds to the eight photometric bands and the fitted mass for each age-metallicity grid point. They are mostly larger than unity, but this should not be interpreted as evidence that SSP models poorly describe the selected GC candidates. A $\chi^2_\nu$ value near unity is expected only if the adopted photometric uncertainties include all relevant sources of scatter and if the model sets are the exact description of the data. In practice, our $\chi^2$ values are computed using the DAOPHOT uncertainties, while other potential systematic uncertainty sources such as SSP model choice, distance, background subtraction, contamination, and the discrete model grid are not included. 
We note that this interpretation is consistent with previous stellar-population fitting studies of GCs. For example, \cite{high_chi_square_value} performed SED fitting for M31 GC S312 and found $\chi^2_\nu$ values of 18.23 and 9.71 for two different Padova model sets, despite obtaining physically meaningful age estimates. Similarly, \cite{high_chi_square_explanation} found average $\chi^2_\nu$ values above unity, and attributed this to unaccounted errors and/or SSP model uncertainties. Other photometric SED fitting studies explicitly include additional model-associated error terms in the fitting uncertainties, illustrating that the photometric errors alone are not always expected to describe the full scatter \citep{high_chi_square_adopt}. In brief, the $\chi^2_\nu$ values in Table~\ref{tab:gc_photometry} are best interpreted as relative goodness of fit within the adopted model grid, rather than as absolute goodness of fit that includes the full uncertainty budget.

\subsection{Mass to Light Ratio}\label{sec:ML_ratio}

The mass-to-light (M/L) ratio serves as a useful diagnostic for understanding the evolution and dynamics of GCs, as well as their stellar population characteristics. 

\begin{figure*}[!htbp]
    \centering
    \includegraphics[width=0.97\linewidth]{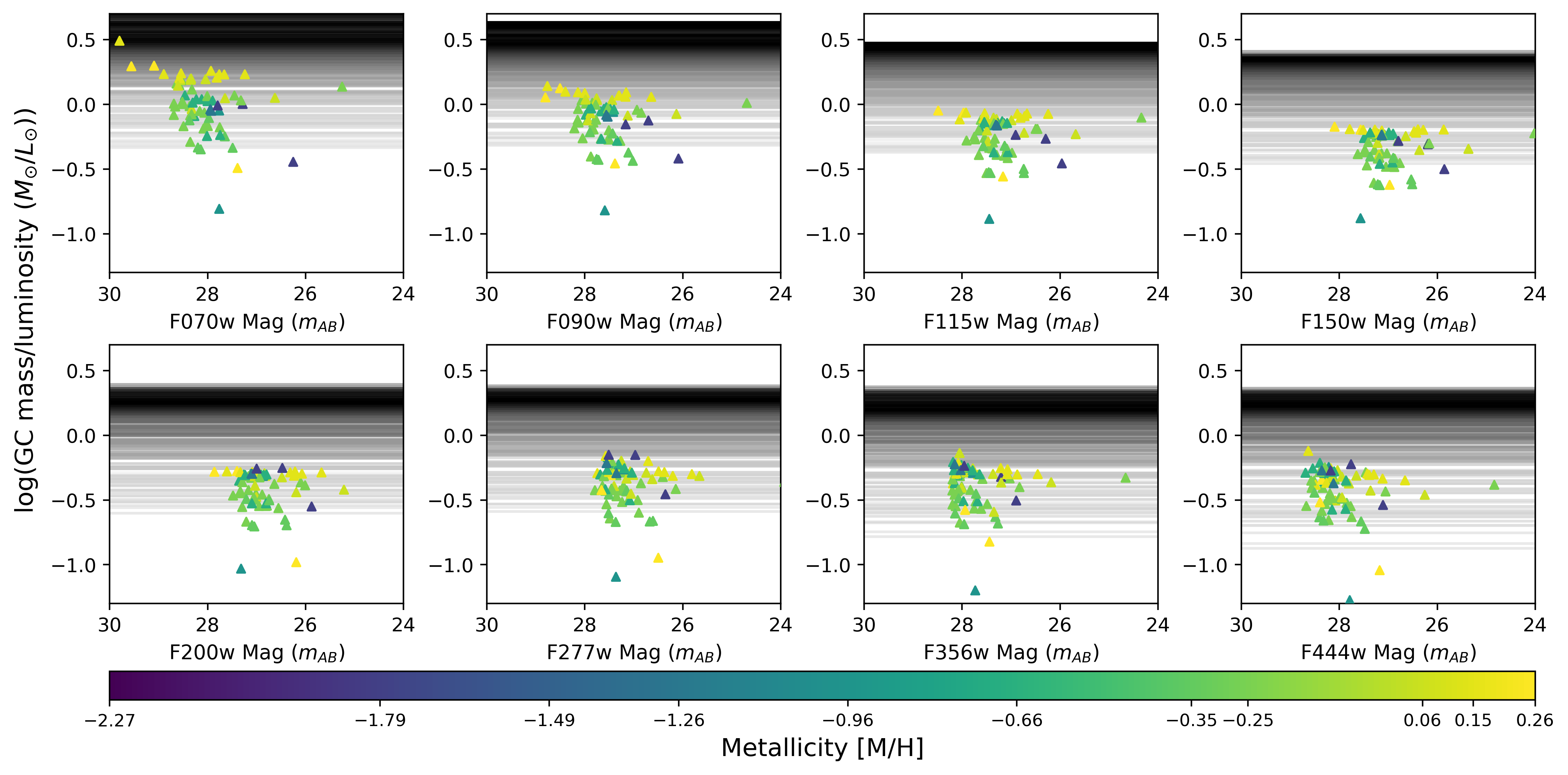}
    \includegraphics[width=0.97\linewidth]{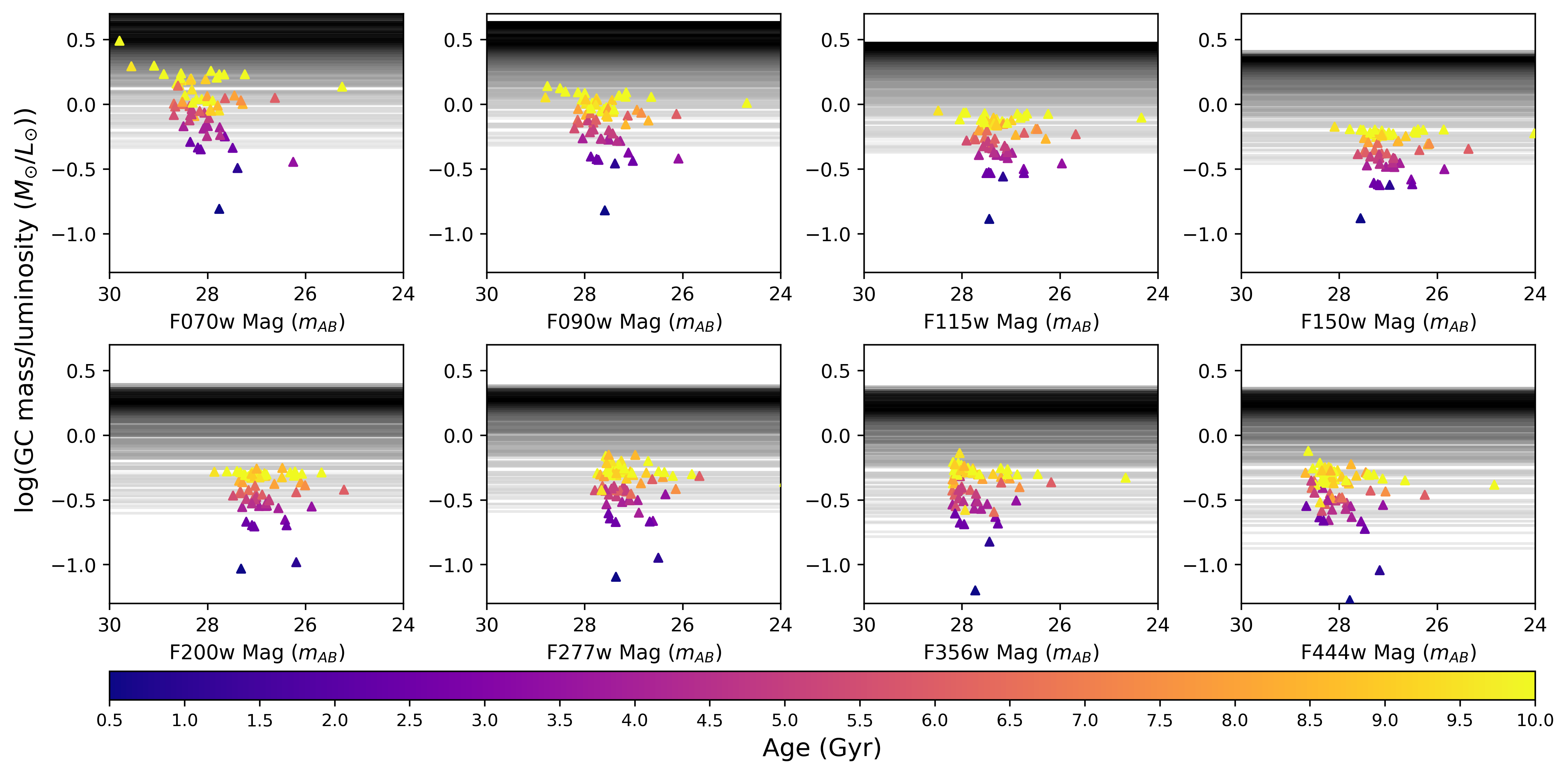}
    \caption{Mass-to-light ratio plots for all filters used in this study. The top half is color-coded by corresponding best-fit metallicities and the bottom half by ages. The grey lines represent PARSEC model predictions for the range of ages and metallicities explored in this work, with darker lines indicating older ages. Since the PARSEC model M/L ratios exhibit no significant dependence on metallicity, no additional color scheme is used to distinguish metallicity in these curves.}
    \label{fig:ML_ratio}
\end{figure*}

Fig.~\ref{fig:ML_ratio} displays the M/L ratios for all filters used in this study, and Table~\ref{tb:ML_ratio} shows the means and scatter of the ratios. The ratios show no strong dependence on magnitude, and they are with modest scatter. The M/L ratios in most filters are notably lower than values reported in previous literature, which primarily focus on shorter wavelengths such as the V-band with M/$\text{L}_V\sim$ 1-3 \citep[e.g.][]{V_band_ratio,ML_v_ratio1,ML_v_ratio2}. This discrepancy likely arises from the fact that M/L ratios tend to be lower at longer wavelengths for these old populations, where the peak emission is longward of $1 \mu$ \citep{lower_ML_ratio}. For a given age, M/L ratio will decrease toward redder filters, reflecting the growing dominance of red‐giant branch stars in the integrated flux. This shows up as a gradual decrease in M/L ratio from F070W through F444W.

\begin{table}
\caption{Mean and scatter (standard deviation) of the mass-to-light ratio across filters in Fig~\ref{fig:ML_ratio}.}
\centering
\label{tb:ML_ratio}
\begin{tabular}{lc}
\hline
Filter & $\langle \log_{10}(M/L)\rangle \pm \sigma$ (dex) \\
\hline
F070W & 0.230 $\pm$ 0.207 \\
F090W & 0.121 $\pm$ 0.179 \\
F115W & -0.008 $\pm$ 0.164 \\
F150W & -0.102 $\pm$ 0.150 \\
F200W & -0.179 $\pm$ 0.156 \\
F277W & -0.143 $\pm$ 0.168 \\
F356W & -0.161 $\pm$ 0.177 \\
F444W & -0.184 $\pm$ 0.179 \\ \hline
\end{tabular}
\end{table}

The color-coding in Fig.~\ref{fig:ML_ratio} follows expected patterns. Metallicity does not appear to significantly affect the M/L ratios, except at shorter wavelengths where variations become relatively more pronounced. In contrast, the effect of age on M/L ratios is more evident: older GC candidates exhibit systematically higher ratios compared to their younger counterparts, consistent with theoretical expectations that older stellar populations retain a larger fraction of their initial mass while gradually losing luminous, high-mass stars.

For comparison, Fig.~\ref{fig:ML_ratio} also shows the M/L ratios predicted by PARSEC stellar population models \citep{PARSEC} over the same age and metallicity range used in this study, plotted as grey lines to provide a theoretical comparison. These lines reproduce the same broad behavior seen in the GC candidates, that M/L ratio increases with age and depends only weakly on metallicity over most of the NIRCam range, and declines toward redder filters at fixed age. In most filters, the PARSEC predictions lie slightly above the observed GC candidate values. This offset is expected, since the PARSEC models give luminosities normalized to the initial stellar mass of the population, rather than the surviving stellar mass.
Therefore, the M/L ratios from PARSEC models in Fig.~\ref{fig:ML_ratio} should be interpreted as approximate upper values for the M/L ratios of real GC systems.

\section{Discussion} \label{sec:discussion}
\subsection{Larger Range of Grid}\label{subsec:larger_grid}
\begin{figure*}[!htbp]
    \centering
    \includegraphics[width=0.99\linewidth]{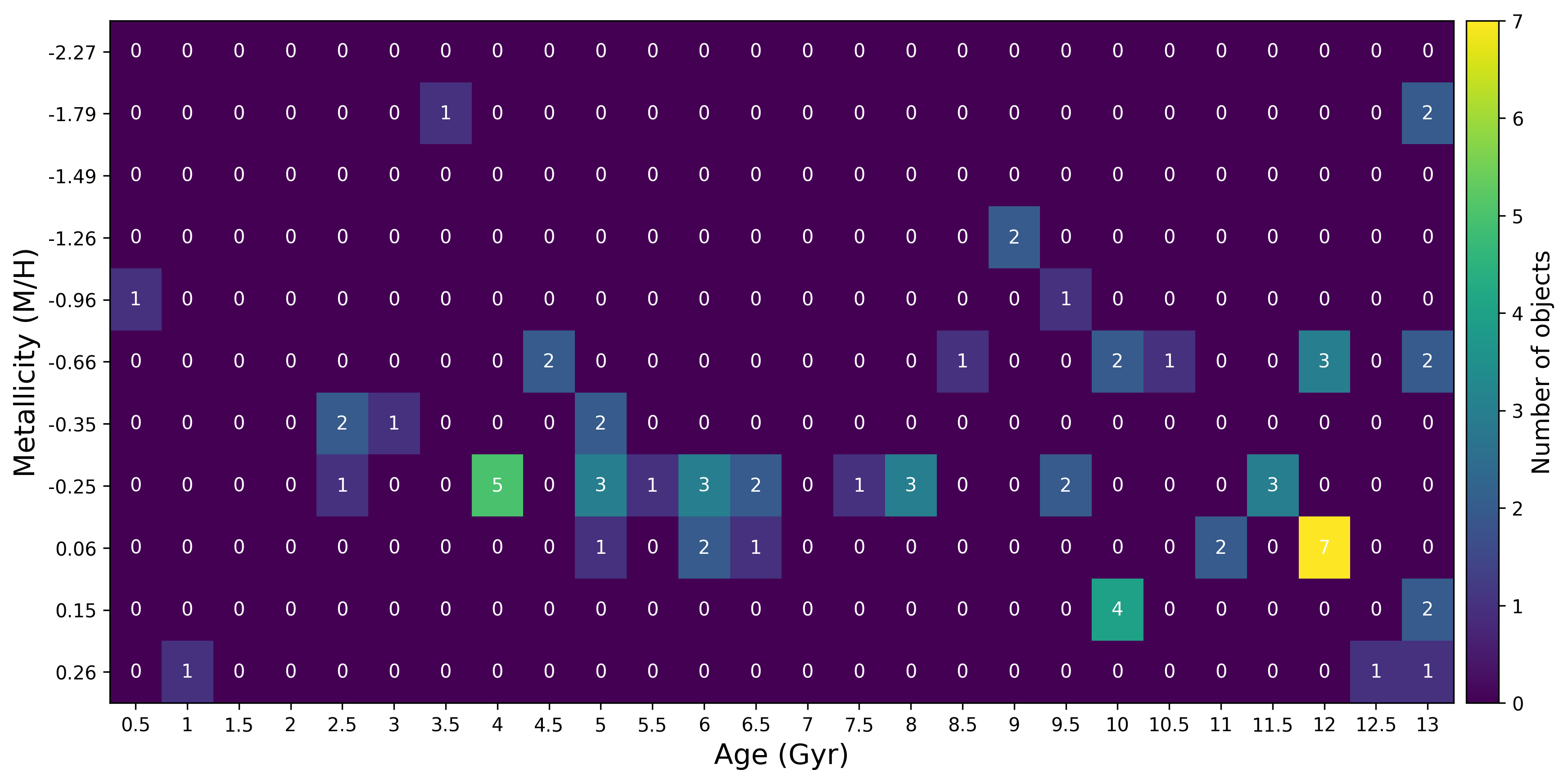}
    \caption{Frequency distribution (heatmap) of best-fit models of a larger grid of ages and metallicities.}
    \label{fig:larger_grid}
\end{figure*}
To evaluate whether our inferred ages and metallicities are artifacts of the original SSP grid sampling or truly reflect the intrinsic SED shapes of the GC candidates, we have refitted all objects on an expanded parameter grid. The new grid spans ages from 0.5 Gyr to 13 Gyr ($\Delta$age$=$0.5 Gyr) and metallicities from –2.27 to +0.26 dex. Although ages of older than 10 Gyr are unphysical at this lookback time (3.5 Gyr), as it would mean GCs older than the universe, we do this strictly as a numerical test to check the potential grid-dependency of SED fitting. Fig.~\ref{fig:larger_grid} shows the resulting heatmap of this larger grid fitting.

A direct comparison between Fig.~\ref{fig:heatmap_upto10} (baseline grid) and Fig.~\ref{fig:larger_grid} (expanded grid) shows that the main features of the age-metallicity grid are preserved. Particularly, the distribution of best fit age and metallicity between the two figures at the early to intermediate ages (between 0.5 and 7 Gyr) are almost perfectly identical, and also there are similar distributions in relatively metal-poor ($\mathrm{[M/H]}\lesssim-0.66$) GC candidates at around 9-10 Gyr.

In the smaller baseline grid, there is a prominent accumulation at the oldest allowed age (10 Gyr) which exhibit a clear bimodality with a metal-rich peak at $[\mathrm{M/H}]=+0.15$ and a metal-poor peak at $[\mathrm{M/H}]=-0.66$. These concentrations at the edge of the model grid could be, in principle, an artifact of the fitting mechanism taken in this paper. However, in the expanded grid, these "old" concentrations remain present, but no longer at the edge of the allowed model grid. These populations are redistributed into slightly older adjacent age bins with the same or similar metallicity. This behavior indicates that the clustering seen in the baseline grid is not driven by effects of the fitting method.
The modest shifts for a subset of GC candidates are discussed further in Section~\ref{subsec:mass_age_degeneracy} and \ref{subsec:age_mh_degeneracy}.
More generally, a large fraction of the GC candidates in Abell 2744 appear to be classically `old', i.e. with ages near the age of the universe at the epoch we are observing them, as they are in the Milky Way \citep[e.g.][]{vandenberg+2013,forbes2020,valenzuela+2024,ying+2025}

\subsection{Dependence on SSP model Choices}\label{subsec:model_dependence}
\begin{figure}[!htbp]
    \centering
    \includegraphics[width=0.98\linewidth]{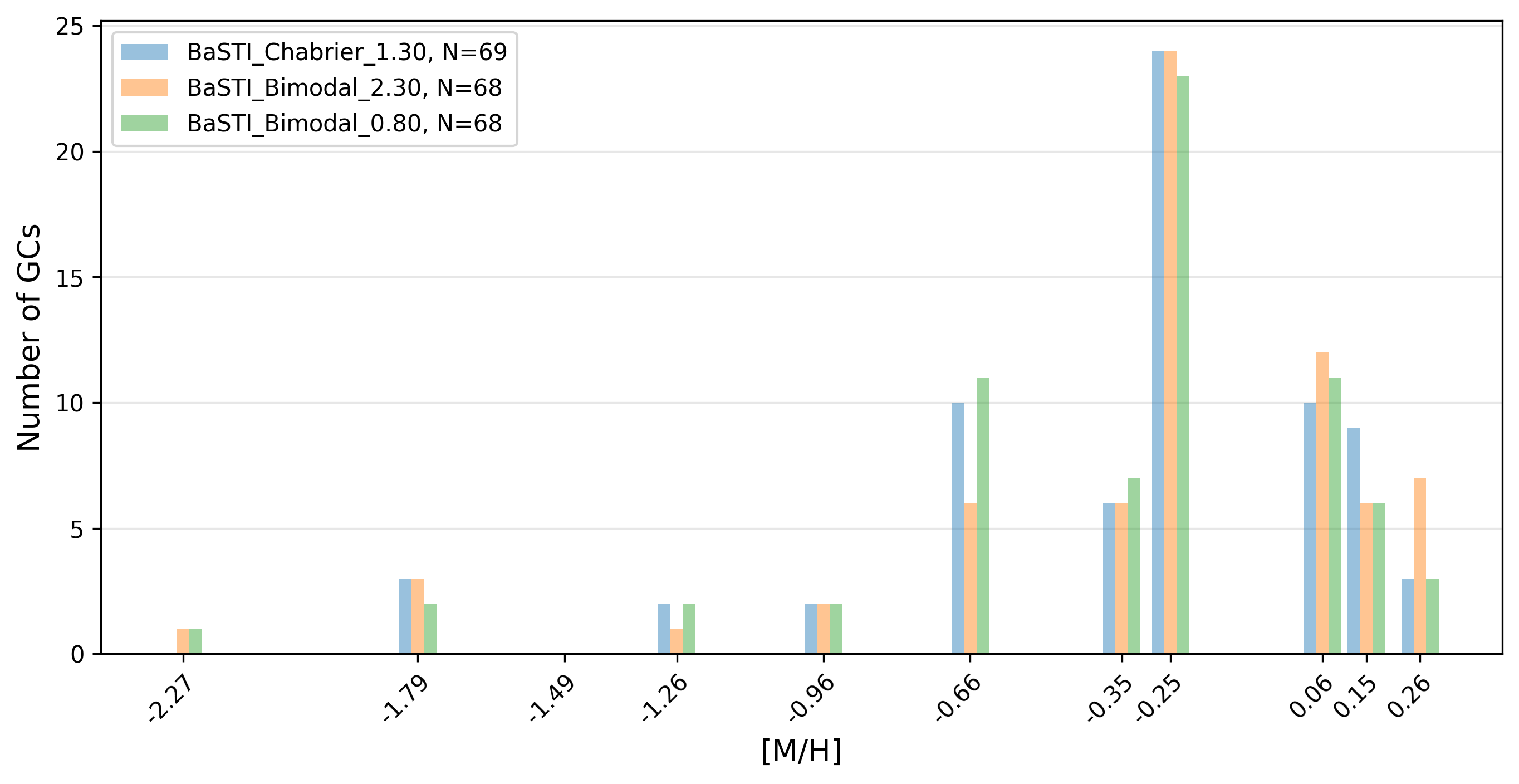}
    \includegraphics[width=0.98\linewidth]{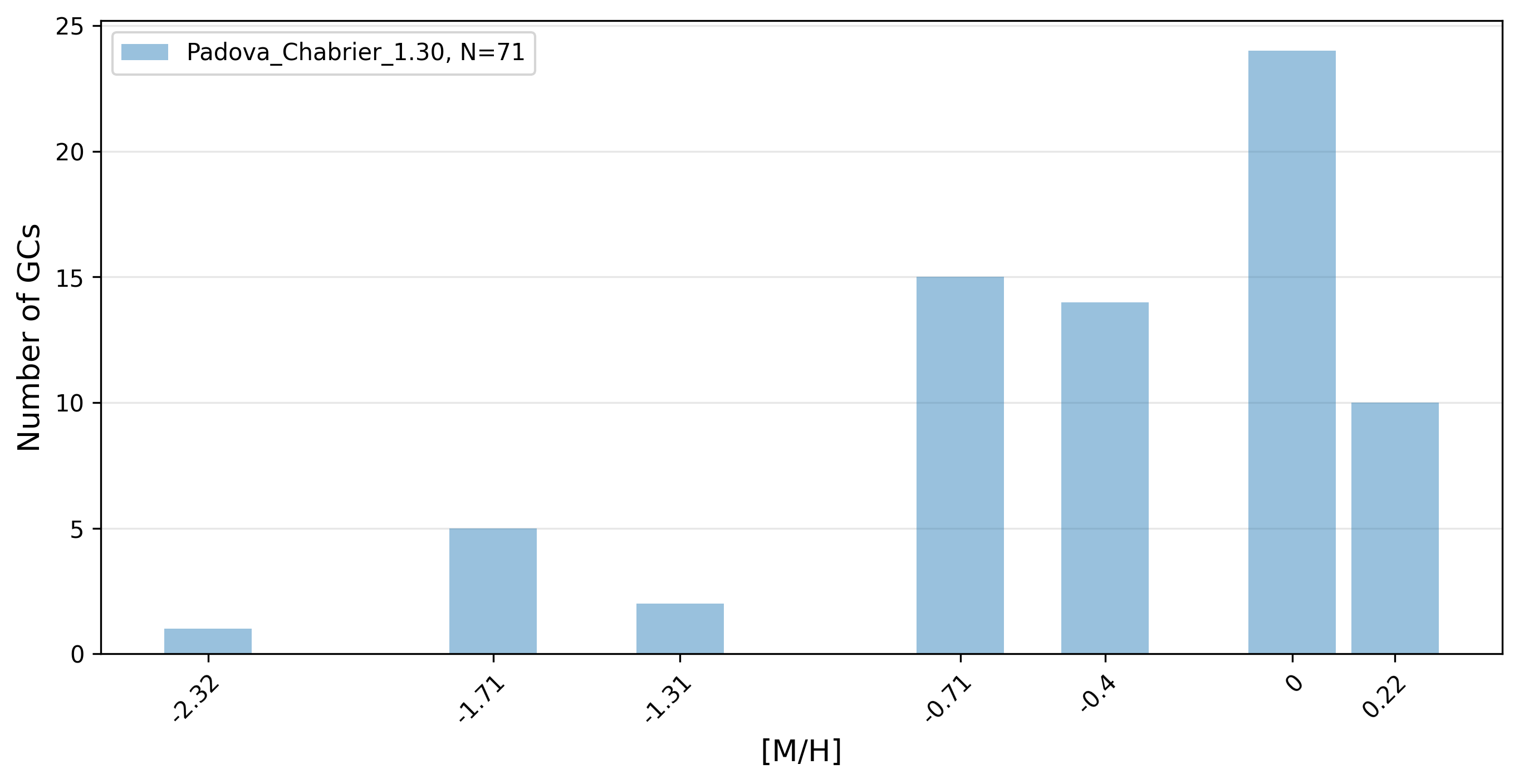}
    \caption{Comparison of best-fit metallicity distributions obtained with different SSP model assumptions. The top panel shows the metallicity distributions with BaSTI model sets with Chabrier IMF (fiducial set) and bimodal IMFs with slopes of 2.30 and 0.80. The bottom panel shows the Padova-isochrone model set with a Chabrier IMF.}
    \label{fig:MH_distribution_multiple_models}
\end{figure}

\begin{figure}[!htbp]
    \centering
    \includegraphics[width=0.98\linewidth]{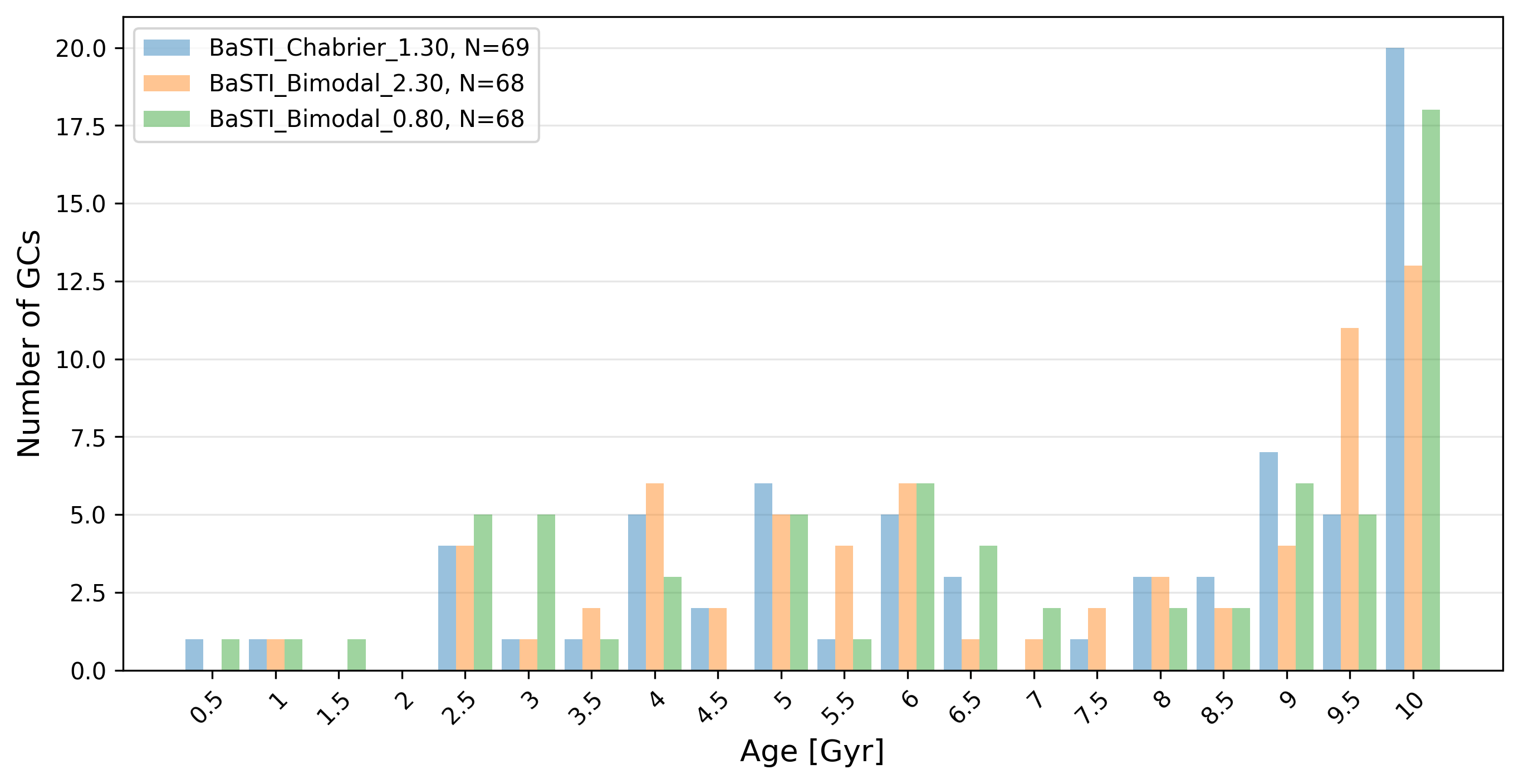}
    \includegraphics[width=0.98\linewidth]{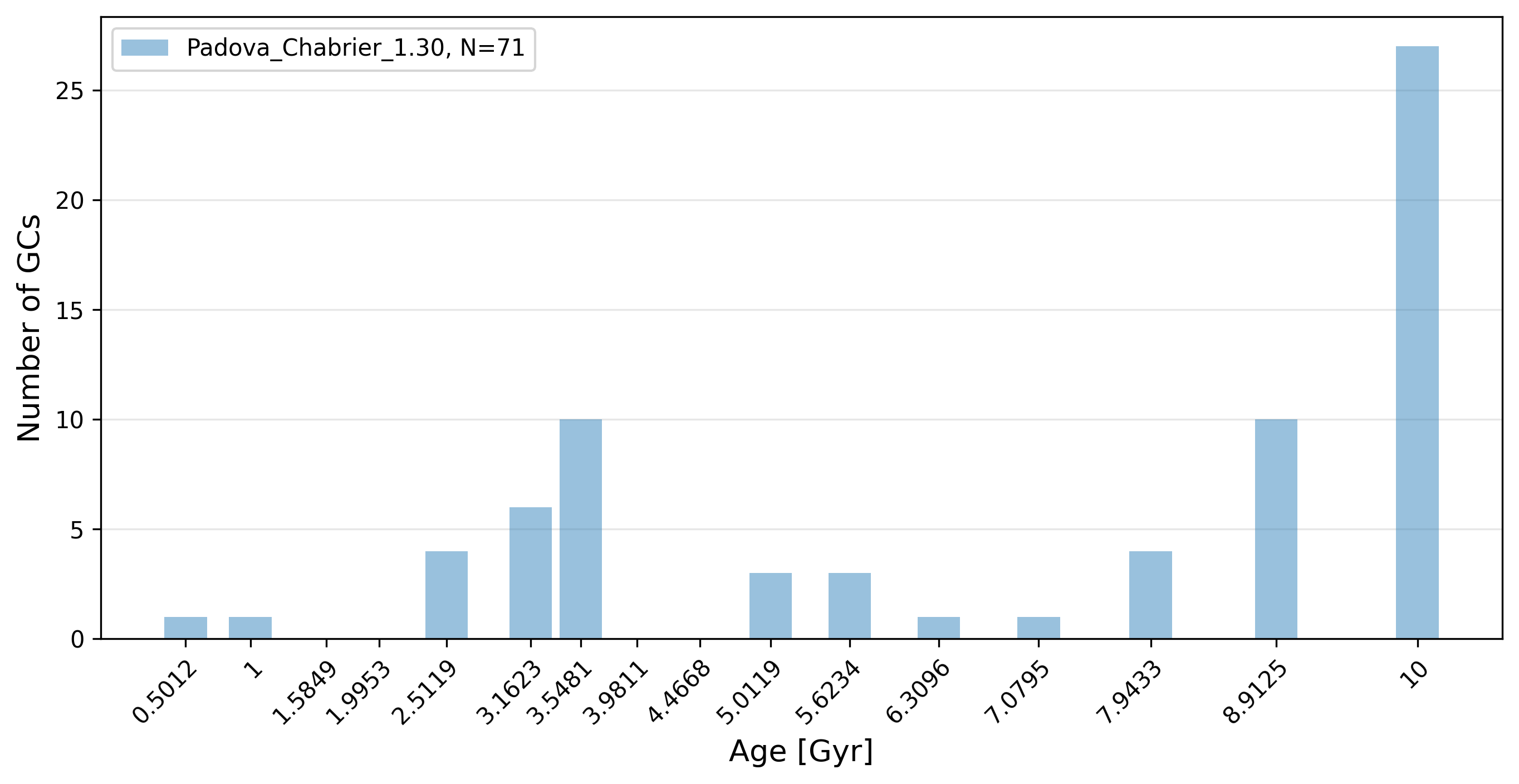}
    \caption{Comparison of best-fit age distributions obtained with different SSP model assumptions. The top panel shows the age distributions with BaSTI model sets with Chabrier IMF (fiducial set) and bimodal IMFs with slopes of 2.30 and 0.80. The bottom panel shows the Padova-isochrone model set with a Chabrier IMF.}
    \label{fig:Age_distribution_multiple_models}
\end{figure}
In addition to testing the effect of the adopted age range, we also examine whether the inferred age and metallicity distributions depend strongly on the choice of SSP model assumptions. The fiducial results in this work are based on the E-MILES BaSTI isochrones with the Chabrier IMF. As a robustness test, we repeat the same fitting procedure using BaSTI models with bimodal IMF slopes of 0.80 and 2.30, as well as Padova isochrone models with the Chabrier IMF. The resulting distributions are shown in  Fig.~\ref{fig:MH_distribution_multiple_models} and \ref{fig:Age_distribution_multiple_models}. The small differences in sample sizes arise because the same selection and fitting criteria are applied separately to each model grid.

BaSTI based model sets show broadly the same distribution in metallicity. In all three cases, the best-fit solutions remain concentrated around [M/H]=-0.25. The age distribution shows somewhat larger model-to-model variations, but all BaSTI grids retain a strong concentration at the old age limit close to $\sim$10 Gyr and a much broader concentration across the intermediate age range.

The Padova-Chabrier solutions produce a slightly different metallicity and age distributions than the BaSTI grids, which is possibly due to the different metallicity binning between the two sets. This could be due to the different bin locations and sizes for this model given by E-MILES, but the overall distributions are still in agreement with each other: higher metal-rich concentration for metallicity and two concentrations at old and intermediate age ranges. These comparisons indicate that the main population-level trends are not driven by a single IMF or isochrone prescriptions, while the detailed object-by-object parameters remain sensitive to SSP model choice.

\subsection{Mass-Age Degeneracy}\label{subsec:mass_age_degeneracy}
\begin{figure}[!htbp]
    \centering
    \includegraphics[width=1.0\linewidth]{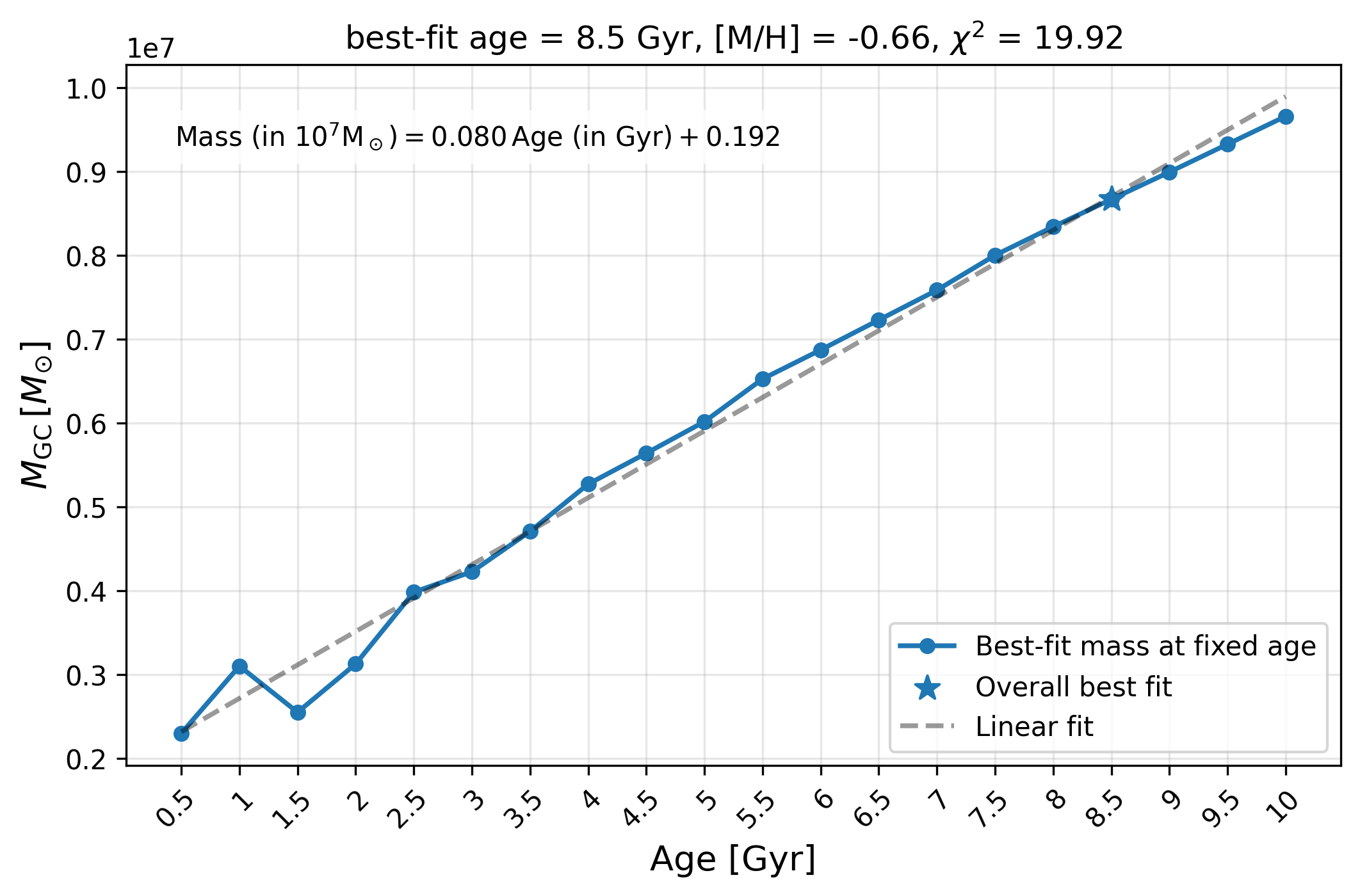}
    \caption{Example of the mass-age trade-off for a representative GC candidate. At each age grid point, the metallicity is fixed to the best-fit value and the mass is refitted to minimize $\chi^2$. The star marks the overall best-fit solution from the full age-metallicity grid. According to the linear-fit, for this particular candidate the mass shifts $\sim8\times10^5 M_{\odot}$ per 1 Gyr. The trade-off between the mass and age is almost linear.}
    \label{fig:mass_age_tradeoff}
\end{figure}

SEDs of SSPs show an intrinsic degeneracy between total stellar mass and age. Both parameters affect the overall normalization of the SED, while age additionally alters its shape only subtly over our wavelength range (see Fig.~\ref{fig:SED_variation}). In our fitting, this degeneracy appears as a correlated behavior—fitting an older SSP model increases the mass estimates, whereas fitting a younger model yields a lower mass estimate. Quantitatively, stepping between age bins (e.g., 8 Gyr $\rightarrow$ 9 Gyr at fixed metallicity) causes mass estimate shifts of $\lesssim$ 10\% per 1 Gyr. Fig.~\ref{fig:mass_age_tradeoff} illustrates this behavior for a representative GC candidate. For this particular candidate, the mass estimate shifts $\sim8\times10^5 M_{\odot}$ per 1 Gyr. The inferred mass increases nearly monotonically with age, showing that older SSP models require larger stellar masses to reproduce the same observed photometric normalization. This trend demonstrates that the fitted mass should not be interpreted independently of the adopted age, because a range of nearby ages can yield similar SED shapes with different mass normalizations.

\begin{figure}[!htbp]
    \centering
    \includegraphics[width=1.0\linewidth]{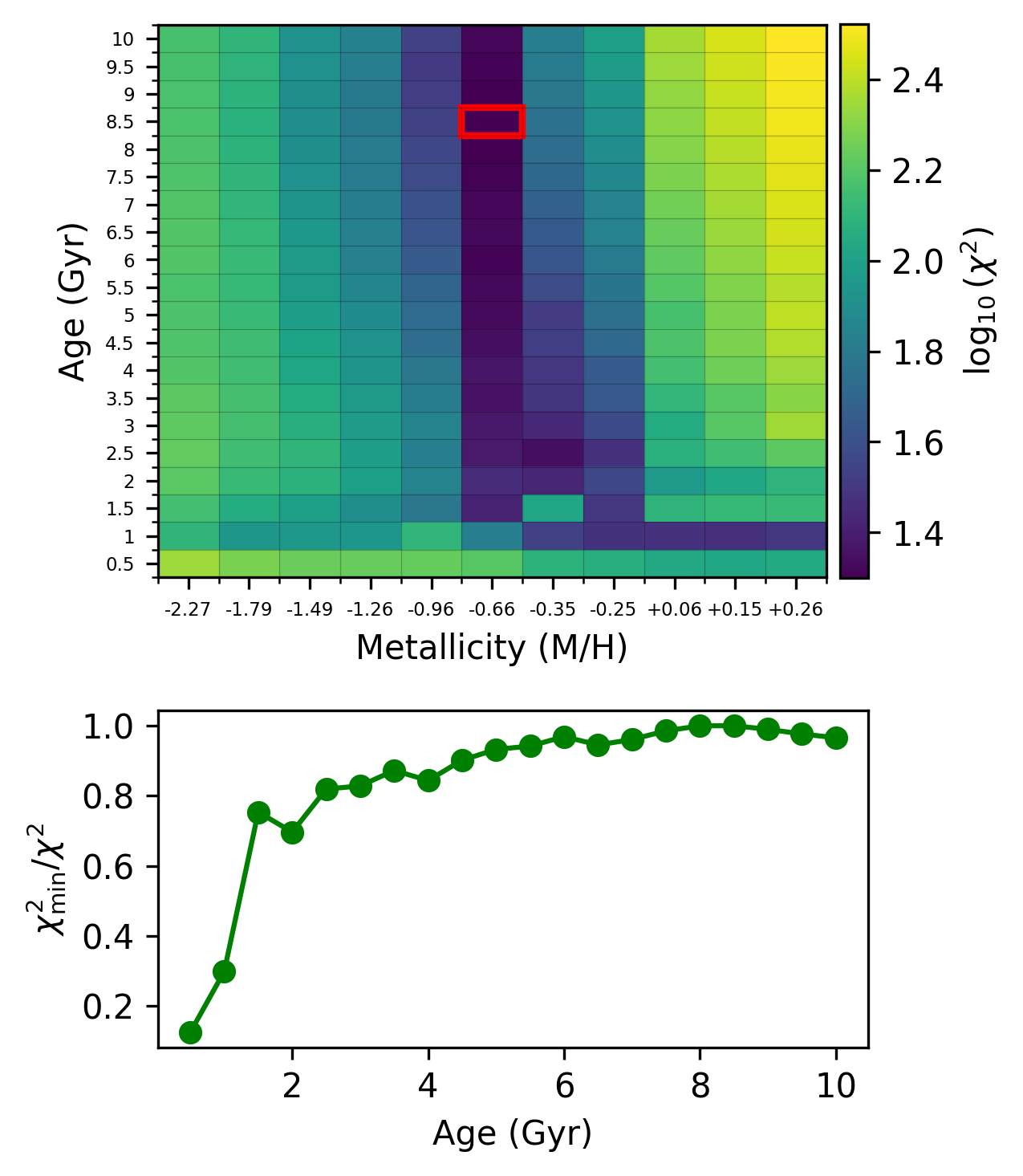}
    \caption{\textit{Top}: Map of the minimum $\chi^2$ across the full age-metallicity model grid (darker colors indicate lower $\chi^2$). The red box marks the global minimum. \textit{Bottom}: Relative fit quality as a function of age at the best-fit metallicity ([M/H]=-0.66 for this object), shown as $\chi^2_{\text{min}}/\chi^2$ where $\chi^2_{\text{min}}$ is the global minimum over the entire grid. The broad, small variation with age illustrates that models near the best-fit age (~8.5 Gyr) provide comparably good fits.}
    \label{fig:age_degeneracy}
\end{figure}

A complementary view of this degeneracy is shown in Fig.~\ref{fig:age_degeneracy}, where we examine the minimum $\chi^2$ values across the full model grid for the same representative candidate. For each point on the model grid (upper plot of the figure), we determine the mass that minimizes $\chi^2$ and then visualize the resulting minimum $\chi^2$ values across the full grid. The heatmap shows a well-defined preference in metallicity, but a relatively broad preference on age. A range of neighboring ages can yield similarly acceptable fits once the mass normalization is adjusted. The best-fit age and metallicity in Fig.~\ref{fig:age_degeneracy} should therefore be interpreted as one location along a relatively flat "valley" in $\chi^2$ with respect to age, rather than as a sharply isolated minimum. Once again, this motivates caution when interpreting the ages of old GCs in broadband SED fitting. 

Because our masses are determined from photometric broadband data, breaking this degeneracy requires an external constraint on mass and/or age. In principle, high-resolution spectroscopy could provide velocity dispersions for individual candidates which would enable dynamical mass estimates that can help to  break the degeneracy, but such work is beyond observational reach at present.

Because broadband SED fitting only weakly constrains GC ages, we perform an additional  robustness test motivated by the expectation that real GCs are predominantly old. We therefore refit the GC candidates with the age fixed to 8 Gyr to assess whether adopting a representative old age alters the inferred best-fit metallicity distribution. The results are presented in Section~\ref{subsec:age_mh_degeneracy}.

\subsection{Age-Metallicity Degeneracy}\label{subsec:age_mh_degeneracy}
\begin{figure}[!htbp]
    \centering
    \includegraphics[width=1.0\linewidth]{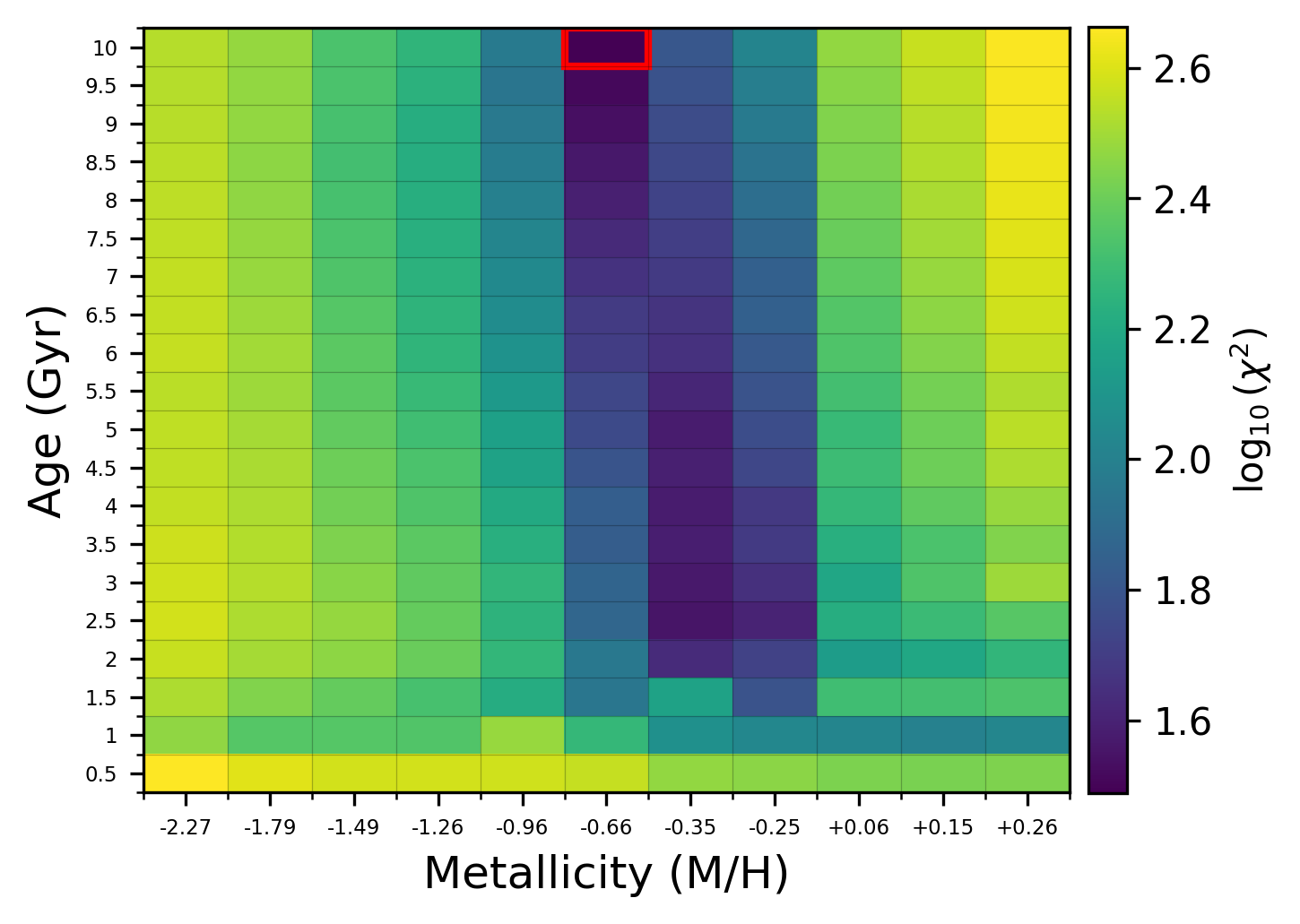}
    \caption{Map of the minimum $\chi^2$ across the full age-metallicity model grid (darker colors indicate lower $\chi^2$). The red box marks the global minimum of this GC candidate. There are two regions with comparable goodness of fit between [M/H]=-0.66 and -0.35, a clear indication of the age-metallicity degeneracy.}
    \label{fig:mh_degeneracy}
\end{figure}

The age-metallicity degeneracy is a well-known problem in GC studies, especially for unresolved or integrated flux data. The degeneracy is due to changes in stellar age and chemical composition, which can produce similar integrated light colors or SED shapes. In broad terms, increasing metallicity reddens the SED through cooler stellar atmospheres, while increasing age also reddens the SED because the main-sequence turnoff
and red-giant branch
shift to lower temperatures \citep{age_mh_degeneracy1}. As a result, a younger, more metal-rich population can resemble an older, more metal-poor population in broadband photometry, especially when the data do not include strongly age-sensitive spectral features.

This degeneracy is a notorious limitation in GC studies. Still, one practical way to reduce this ambiguity is to model the full multi-band SED that extends into the NIR \citep{opt_to_NIR}. Effectively, multi-band photometric SED fitting gives additional leverage that is less degenerate than optical-only color-color space, although this does not eliminate the degeneracy. 

\begin{figure}
    \centering
    \includegraphics[width=0.97\linewidth]{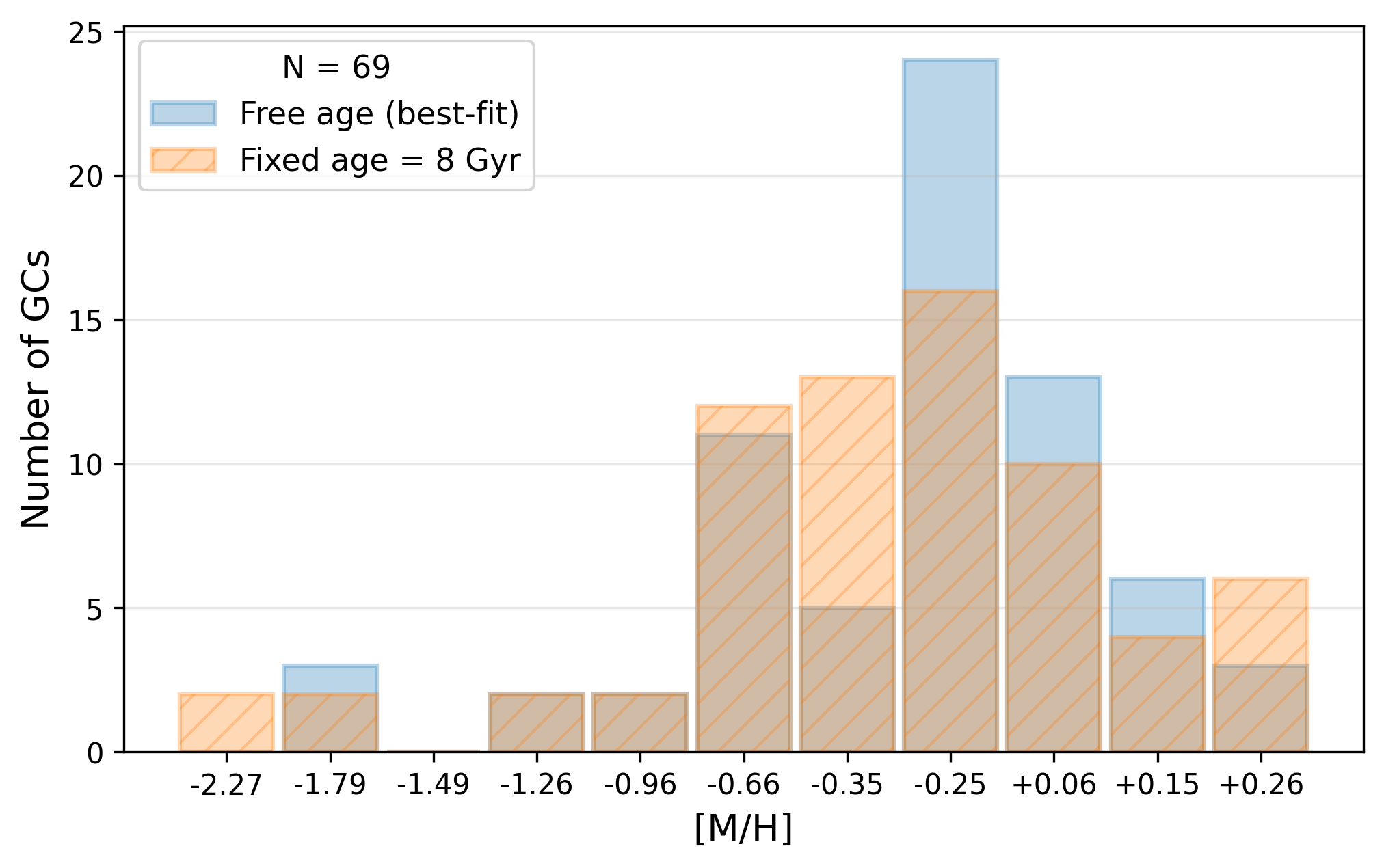}
    \caption{Overlaid best-fit metallicity distributions for the GC candidates. Blue shows the best-fit metallicity values when age is a free parameter, and orange (hatched) shows results when the age is fixed to 8 Gyr. Fixing the age primarily redistributes candidates among nearby metallicity bins.}
    \label{fig:fixed_age}
\end{figure}

In our study, this degeneracy appears in multiple GC candidates directly in the structure of their $\chi^2$ surfaces in the age-metallicity grid. Fig.~\ref{fig:mh_degeneracy} shows that acceptable solutions are not confined to a single sharply isolated minimum, but instead it is a locus of comparably acceptable solutions along which age and metallicity trade off with only a small change in goodness of fit. Fig.~\ref{fig:fixed_age} shows the practical impact of this degeneracy on the inferred metallicity distribution by comparing the best-fit metallicity values from the baseline fits (with age as a free parameter) to those obtained when the age is fixed to 8 Gyr (motivated in Section~\ref{subsec:mass_age_degeneracy}). The two distributions remain broadly similar, and fixing the age does not result in a qualitatively distinct distribution. Instead, it redistributes a subset of candidates among nearby metallicity bins. This behavior is expected from the minimum $\chi^2$ structure in Fig.~\ref{fig:mh_degeneracy}, where age-metallicity tradeoffs yield comparably acceptable solutions. This shows that the primary effect of imposing an old-age prior is to move the object's best-fit metallicity to adjacent nodes rather than to completely overturn the metallicity preference of the object.

This behavior can also explain the changes in best-fit values when the model grid is widened in Section~\ref{subsec:larger_grid}. The extended grid effectively adds nearby "escape routes" in parameter space, so the formal minimum can shift from one node to an adjacent one even when the underlying SED is nearly unchanged. These shifts in metallicity only occur in nearby metallicity regions with typical differences in metallicity $\Delta\mathrm{[M/H]}\lesssim0.3$.

\subsection{LWC Dimming}
During the fitting process, magnitudes in NIRCam LWC are often observed to be dimmer than their best-fit SSP models. This behavior is shown in Fig.~\ref{fig:Good_fits_mosaic}, in each of the F277W, F356W, and F444W filters, many of the candidates' observed magnitudes 
are fainter than
the model predictions.
Such a consistent dimming cannot be attributed to random photometric errors alone; rather, it suggests a systematic effect either in the data reduction or in the assumed model SEDs.

Several factors may contribute to the systematic LWC under‐luminosity. First, treating each candidate as a pure SSP may omit key components. Many SSP libraries vary in their treatment of nebular continuum, IMF prescription, and dust emission, AGB star contribution, and many more, all of which shape the overall SED. Especially, for the IMF prescription, the real GCs might have different IMF compared to the assumed model. Furthermore, the mass estimates are in the mass regime where there is overlap between GCs and UCDs (for example, $\geq 2\times10^6M_{\odot}$ by \cite{GC_UCD_1}.  Their IMFs may differ, and UCDs may hold more complex stellar populations \cite{GC_UCD_2}). The E‑MILES models used here may therefore lack some of the features present in our targets. Second, some candidates may appear artificially faint in the LWC if the local background is over-subtracted during PSF fitting photometry. In PSF fitting, the background level is estimated locally, and small systematic over- or under-subtractions can arise depending on the exact complexity of the local background light distribution. This effect is expected to be more significant in F356W and F444W because their lower spatial resolution compared to SWC increases blending and crowding, thereby amplifying uncertainties.

\section{Conclusion} \label{sec:conclusion}
We summarize the key findings of this study as follows:
\begin{enumerate}
    \item By combining eight NIRCam filters (F070W to F444W) and E‑MILES SSP models, we performed $\chi^2$ minimization SED fitting for 69 GC candidates in Abell 2744 (z = 0.308). Photometric uncertainties were propagated with Monte Carlo resampling of the observed magnitudes, providing formal uncertainty estimates on age, metallicity, and stellar mass.
    \item The inferred stellar masses cluster tightly around $10^7M_{\odot}$, consistent with the high‑mass tail of local GC populations, likely overlapping with some UCD-like compact stellar systems. Future extension of the analysis to fainter candidates will allow the same SED-fitting framework to probe the more typical GC mass regime.
    \item There are peaks in the best-fit age-metallicity distributions: [M/H]=-0.25 at intermediate age (4-6.5 Gyr), [M/H]=-0.66 at 10 Gyr and [M/H]=+0.15 at 10 Gyr. This result is plausible from the known age and metallicity ranges of GCs in Milky Way, but is not certain given the insensitivity of the SED fitting due to degeneracies. At face value, however, our results are suggestive of two major epochs of GC/UCD formation, one around 3 -- 6 Gyr and another around 8 -- 10 Gyr.
    \item Across the NIRCam filters, the derived M/L ratios increase systematically with age, show only weak dependence on metallicity, and generally decline toward redder filters for a fixed age. These trends are consistent with the theoretical stellar-population predictions, which older populations have higher M/L ratios and longer wavelength light is increasingly dominated by red giant branch stars.
    \item There is mass-age degeneracy in this fitting method that requires spectroscopic or dynamical priors for better estimates. Quantitatively, there is $\sim10\%$ shift in mass per 1 Gyr age step.
    \item For some GC candidates, the SED fitting results are not certain due to age-metallicity degeneracy, yet their metallicities are mostly constrained within $\Delta\mathrm{[M/H]}\lesssim0.3$.
\end{enumerate}
These results underscore JWST’s power to characterize GCs at intermediate redshift and point toward future work. The metallicity distributions especially are the most robust results of these studies. Looking ahead, there will be JWST observations containing GC populations that allow us to extend and refine this SED‐fitting approach across diverse cluster environments and a large range of lookback times. With these datasets becoming increasingly common, our approach will provide progressively tighter constraints on GC formation histories and evolution, marking this moment as the beginning of a new era in extragalactic stellar‐population studies.

\appendix

\FloatBarrier
\section{Table}\label{app:tables}
\startlongtable
\input{bestfit_with_mc_errors}

\bibliography{main}{}
\bibliographystyle{aasjournalv7}

\end{document}

%% file: bestfit_with_mc_errors.tex
\begin{deluxetable*}{ccccccccccccc}
\rotate
\tabletypesize{\scriptsize}
\tablewidth{0pt}
\tablecaption{Observed magnitudes and best-fit parameters for the 69 GC candidates. Uncertainties on age, metallicity, and stellar mass are estimated from Monte Carlo photometric error propagation using the 16th and 84th percentiles.\label{tab:gc_photometry}}
\tablehead{
\colhead{ID} &
\colhead{Age (Gyr)} &
\colhead{[M/H]} &
\colhead{$\log(M_\star/M_\odot)$} &
\colhead{$\chi^2_\nu$} &
\colhead{F070W} &
\colhead{F090W} &
\colhead{F115W} &
\colhead{F150W} &
\colhead{F200W} &
\colhead{F277W} &
\colhead{F356W} &
\colhead{F444W}
}
\startdata
1 & $3.5^{+0.0}_{-1.0}$ & $-1.79^{+0.00}_{-0.48}$ & $7.31^{+0.01}_{-0.09}$ & 5.63 & 26.25 & 26.09 & 25.96 & 25.86 & 25.88 & 26.35 & 26.89 & 27.11 \\
2 & $4.0^{+6.0}_{-1.5}$ & $-0.35^{+0.10}_{-0.31}$ & $7.11^{+0.24}_{-0.14}$ & 6.31 & 27.65 & 27.11 & 26.74 & 26.53 & 26.42 & 26.61 & 27.32 & 27.56 \\
3 & $9.0^{+1.0}_{-2.0}$ & $0.06^{+0.09}_{-0.00}$ & $7.20^{+0.03}_{-0.07}$ & 2.54 & 28.35 & 27.75 & 27.30 & 26.97 & 26.82 & 27.07 & 27.78 & 27.80 \\
4 & $5.0^{+1.0}_{-0.0}$ & $-0.25^{+0.00}_{-0.00}$ & $6.91^{+0.05}_{-0.01}$ & 6.98 & 28.36 & 27.87 & 27.45 & 27.21 & 27.08 & 27.35 & 28.10 & 28.54 \\
5 & $9.5^{+0.5}_{-2.5}$ & $0.06^{+0.00}_{-0.00}$ & $7.22^{+0.02}_{-0.08}$ & 6.96 & 28.34 & 27.76 & 27.30 & 26.98 & 26.84 & 27.26 & 28.05 & 28.04 \\
6 & $8.0^{+1.5}_{-2.0}$ & $-0.25^{+0.00}_{-0.00}$ & $7.44^{+0.05}_{-0.08}$ & 5.59 & 27.45 & 26.94 & 26.50 & 26.19 & 26.10 & 26.41 & 27.03 & 27.46 \\
7 & $9.5^{+0.5}_{-1.1}$ & $-0.25^{+0.00}_{-0.00}$ & $7.07^{+0.01}_{-0.03}$ & 7.15 & 28.58 & 27.98 & 27.52 & 27.25 & 27.11 & 27.36 & 28.04 & 28.22 \\
8 & $4.0^{+1.0}_{-1.5}$ & $-0.25^{+0.00}_{-0.10}$ & $6.97^{+0.06}_{-0.13}$ & 11.26 & 28.02 & 27.52 & 27.07 & 26.88 & 26.79 & 27.11 & 27.72 & 27.75 \\
9 & $9.0^{+1.0}_{-1.6}$ & $0.06^{+0.09}_{-0.00}$ & $7.33^{+0.03}_{-0.06}$ & 10.49 & 28.04 & 27.44 & 26.99 & 26.64 & 26.48 & 26.74 & 27.37 & 27.65 \\
10 & $4.0^{+1.0}_{-1.0}$ & $-0.25^{+0.00}_{-0.10}$ & $6.96^{+0.06}_{-0.09}$ & 5.15 & 28.00 & 27.52 & 27.14 & 26.94 & 26.81 & 26.90 & 27.61 & 28.34 \\
11 & $10.0^{+0.0}_{-7.5}$ & $-0.66^{+0.31}_{-0.00}$ & $7.13^{+0.00}_{-0.38}$ & 5.75 & 28.12 & 27.61 & 27.27 & 27.08 & 26.99 & 27.53 & 28.01 & 28.40 \\
12 & $3.0^{+2.0}_{-0.5}$ & $-0.35^{+0.10}_{-0.00}$ & $6.70^{+0.16}_{-0.04}$ & 6.21 & 28.35 & 27.88 & 27.50 & 27.30 & 27.21 & 27.49 & 28.05 & 28.67 \\
13 & $10.0^{+0.0}_{-0.5}$ & $-0.25^{+0.00}_{-0.00}$ & $7.12^{+0.01}_{-0.01}$ & 10.24 & 28.53 & 27.85 & 27.39 & 27.12 & 27.02 & 27.27 & 27.95 & 28.09 \\
14 & $5.0^{+1.0}_{-1.0}$ & $-0.25^{+0.00}_{-0.00}$ & $7.04^{+0.05}_{-0.06}$ & 8.69 & 28.07 & 27.51 & 27.09 & 26.87 & 26.75 & 27.21 & 28.04 & 27.85 \\
15 & $9.0^{+1.0}_{-2.0}$ & $0.06^{+0.09}_{-0.00}$ & $7.13^{+0.03}_{-0.07}$ & 9.02 & 28.56 & 27.98 & 27.47 & 27.21 & 27.05 & 27.15 & 27.92 & 28.22 \\
16 & $6.5^{+3.0}_{-0.5}$ & $-0.25^{+0.00}_{-0.00}$ & $6.95^{+0.10}_{-0.03}$ & 4.53 & 28.49 & 27.95 & 27.54 & 27.28 & 27.16 & 27.37 & 28.04 & 28.51 \\
17 & $8.5^{+0.5}_{-0.0}$ & $-1.79^{+0.30}_{-0.00}$ & $7.37^{+0.01}_{-0.01}$ & 9.96 & 27.28 & 26.70 & 26.29 & 26.19 & 26.47 & 26.97 & 27.20 & 27.76 \\
18 & $10.0^{+0.0}_{-0.0}$ & $0.15^{+0.00}_{-0.00}$ & $6.93^{+0.00}_{-0.00}$ & 19.75 & 29.80 & 28.77 & 28.05 & 27.79 & 27.61 & 27.74 & 28.19 & 28.49 \\
19 & $0.5^{+0.0}_{-0.0}$ & $-0.96^{+0.00}_{-1.31}$ & $6.28^{+0.01}_{-0.02}$ & 8.08 & 27.76 & 27.59 & 27.45 & 27.57 & 27.32 & 27.36 & 27.73 & 27.79 \\
20 & $6.0^{+0.5}_{-1.0}$ & $-0.25^{+0.00}_{-0.00}$ & $6.98^{+0.02}_{-0.05}$ & 9.07 & 28.37 & 27.83 & 27.37 & 27.14 & 27.01 & 27.28 & 28.01 & 28.19 \\
21 & $10.0^{+0.0}_{-5.0}$ & $-0.66^{+0.31}_{-0.00}$ & $7.02^{+0.00}_{-0.17}$ & 4.39 & 28.46 & 27.90 & 27.53 & 27.36 & 27.26 & 27.55 & 28.09 & 28.31 \\
22 & $10.0^{+0.0}_{-0.0}$ & $0.26^{+0.00}_{-0.11}$ & $7.02^{+0.00}_{-0.02}$ & 3.70 & 29.09 & 28.51 & 27.97 & 27.57 & 27.40 & 27.47 & 28.16 & 28.35 \\
23 & $10.0^{+0.0}_{-1.0}$ & $0.15^{+0.00}_{-0.09}$ & $7.47^{+0.01}_{-0.03}$ & 4.26 & 27.76 & 27.17 & 26.72 & 26.39 & 26.21 & 26.38 & 27.07 & 27.29 \\
24 & $10.0^{+0.0}_{-1.5}$ & $-0.66^{+0.00}_{-0.00}$ & $7.09^{+0.01}_{-0.05}$ & 4.55 & 28.23 & 27.68 & 27.34 & 27.18 & 27.12 & 27.44 & 28.18 & 28.31 \\
25 & $7.5^{+2.0}_{-1.5}$ & $-0.25^{+0.00}_{-0.00}$ & $6.92^{+0.07}_{-0.08}$ & 1.82 & 28.69 & 28.15 & 27.75 & 27.48 & 27.33 & 27.64 & 28.20 & 28.57 \\
26 & $9.0^{+1.0}_{-3.0}$ & $0.06^{+0.09}_{-0.00}$ & $7.21^{+0.03}_{-0.11}$ & 3.36 & 28.34 & 27.76 & 27.36 & 26.99 & 26.82 & 27.02 & 27.67 & 27.95 \\
27 & $4.0^{+1.0}_{-1.0}$ & $-0.25^{+0.00}_{-0.00}$ & $6.76^{+0.06}_{-0.09}$ & 2.67 & 28.49 & 28.05 & 27.66 & 27.44 & 27.29 & 27.56 & 28.20 & 28.22 \\
28 & $10.0^{+0.0}_{-0.0}$ & $0.15^{+0.00}_{-0.00}$ & $7.52^{+0.00}_{-0.00}$ & 2.79 & 27.66 & 27.16 & 26.67 & 26.29 & 26.07 & 26.20 & 26.87 & 27.12 \\
29 & $9.5^{+0.5}_{-2.5}$ & $-0.25^{+0.31}_{-0.00}$ & $7.05^{+0.01}_{-0.05}$ & 4.42 & 28.63 & 28.07 & 27.62 & 27.33 & 27.16 & 27.48 & 28.12 & 28.40 \\
30 & $3.5^{+5.0}_{-1.0}$ & $-0.66^{+0.31}_{-0.00}$ & $6.67^{+0.26}_{-0.08}$ & 2.85 & 28.27 & 27.98 & 27.61 & 27.50 & 27.36 & 27.69 & 28.15 & 28.69 \\
31 & $6.5^{+3.0}_{-0.5}$ & $-0.25^{+0.00}_{-0.00}$ & $7.41^{+0.10}_{-0.03}$ & 3.88 & 27.32 & 26.85 & 26.46 & 26.16 & 26.01 & 26.14 & 26.83 & 27.06 \\
32 & $10.0^{+0.0}_{-0.5}$ & $0.15^{+0.00}_{-0.00}$ & $7.69^{+0.00}_{-0.01}$ & 2.42 & 27.24 & 26.64 & 26.24 & 25.87 & 25.67 & 25.81 & 26.45 & 26.66 \\
33 & $6.0^{+2.0}_{-1.0}$ & $-0.25^{+0.00}_{-0.00}$ & $6.86^{+0.09}_{-0.05}$ & 1.63 & 28.66 & 28.13 & 27.73 & 27.46 & 27.32 & 27.58 & 28.15 & 28.29 \\
34 & $10.0^{+0.0}_{-0.0}$ & $0.15^{+0.00}_{-0.00}$ & $7.18^{+0.00}_{-0.00}$ & 3.60 & 28.54 & 28.00 & 27.54 & 27.13 & 26.94 & 27.18 & 27.79 & 28.18 \\
35 & $9.0^{+1.0}_{-2.0}$ & $0.06^{+0.09}_{-0.00}$ & $7.11^{+0.03}_{-0.07}$ & 9.99 & 28.56 & 27.99 & 27.50 & 27.23 & 27.04 & 27.36 & 27.97 & 28.15 \\
36 & $9.0^{+0.5}_{-5.0}$ & $-0.96^{+0.30}_{-0.30}$ & $7.14^{+0.02}_{-0.23}$ & 4.97 & 27.76 & 27.42 & 27.14 & 26.98 & 26.87 & 27.21 & 27.78 & 28.22 \\
37 & $2.5^{+2.5}_{-0.0}$ & $-0.35^{+0.00}_{-0.00}$ & $6.70^{+0.20}_{-0.01}$ & 2.40 & 28.20 & 27.76 & 27.46 & 27.21 & 27.09 & 27.37 & 27.96 & 28.41 \\
38 & $9.5^{+0.5}_{-3.5}$ & $-0.25^{+0.31}_{-0.00}$ & $8.38^{+0.01}_{-0.10}$ & 4.72 & 25.25 & 24.69 & 24.34 & 24.02 & 23.85 & 23.93 & 24.66 & 24.83 \\
39 & $5.0^{+4.5}_{-1.5}$ & $0.06^{+0.00}_{-0.31}$ & $6.97^{+0.13}_{-0.10}$ & 5.21 & 28.32 & 27.93 & 27.47 & 27.19 & 27.01 & 27.40 & 28.01 & 28.08 \\
40 & $6.5^{+3.0}_{-1.5}$ & $-0.25^{+0.00}_{-0.00}$ & $6.89^{+0.11}_{-0.07}$ & 6.07 & 28.52 & 28.14 & 27.66 & 27.41 & 27.29 & 27.61 & 28.19 & 28.15 \\
41 & $10.0^{+0.0}_{-7.5}$ & $-0.66^{+0.31}_{-0.00}$ & $7.18^{+0.00}_{-0.39}$ & 6.41 & 28.01 & 27.48 & 27.18 & 26.99 & 26.85 & 27.19 & 27.78 & 28.10 \\
42 & $3.0^{+7.0}_{-0.5}$ & $-0.35^{+0.00}_{-0.31}$ & $7.03^{+0.34}_{-0.05}$ & 3.59 & 27.49 & 27.02 & 26.74 & 26.51 & 26.39 & 26.67 & 27.27 & 27.48 \\
43 & $10.0^{+0.0}_{-7.5}$ & $-0.66^{+0.31}_{-0.00}$ & $7.16^{+0.01}_{-0.38}$ & 6.98 & 28.00 & 27.56 & 27.20 & 26.99 & 26.90 & 27.31 & 27.88 & 28.14 \\
44 & $4.5^{+2.5}_{-1.0}$ & $-0.25^{+0.00}_{-0.10}$ & $7.08^{+0.13}_{-0.07}$ & 3.15 & 27.75 & 27.28 & 26.98 & 26.77 & 26.55 & 26.92 & 27.49 & 27.77 \\
45 & $6.5^{+3.0}_{-1.0}$ & $-0.25^{+0.00}_{-0.00}$ & $7.14^{+0.11}_{-0.03}$ & 5.96 & 28.01 & 27.53 & 27.16 & 26.81 & 26.64 & 26.85 & 27.58 & 27.85 \\
46 & $10.0^{+0.0}_{-0.0}$ & $0.15^{+0.00}_{-0.00}$ & $7.03^{+0.00}_{-0.00}$ & 2.51 & 28.89 & 28.40 & 27.92 & 27.51 & 27.33 & 27.51 & 28.12 & 28.26 \\
47 & $3.5^{+1.0}_{-1.0}$ & $-0.25^{+0.00}_{-0.00}$ & $6.86^{+0.08}_{-0.08}$ & 2.31 & 28.08 & 27.67 & 27.30 & 27.05 & 26.95 & 27.24 & 27.75 & 28.24 \\
48 & $10.0^{+0.0}_{-2.0}$ & $-0.66^{+0.00}_{-0.00}$ & $7.21^{+0.00}_{-0.06}$ & 2.66 & 27.90 & 27.40 & 27.08 & 26.90 & 26.80 & 27.04 & 27.64 & 27.86 \\
49 & $6.0^{+2.0}_{-1.0}$ & $-0.25^{+0.00}_{-0.00}$ & $7.02^{+0.08}_{-0.06}$ & 3.63 & 28.16 & 27.77 & 27.33 & 27.03 & 26.89 & 27.16 & 27.83 & 28.00 \\
50 & $8.0^{+1.5}_{-3.5}$ & $-0.66^{+0.00}_{-0.30}$ & $7.14^{+0.06}_{-0.17}$ & 1.82 & 27.74 & 27.34 & 27.07 & 26.91 & 26.81 & 27.08 & 27.68 & 27.87 \\
51 & $5.0^{+1.0}_{-1.0}$ & $-0.25^{+0.00}_{-0.00}$ & $6.92^{+0.05}_{-0.06}$ & 7.71 & 28.33 & 27.82 & 27.39 & 27.18 & 27.04 & 27.46 & 28.07 & 28.57 \\
52 & $5.0^{+3.5}_{-3.0}$ & $-0.25^{+0.00}_{-0.10}$ & $6.76^{+0.14}_{-0.31}$ & 0.28 & 28.69 & 28.21 & 27.91 & 27.62 & 27.48 & 27.79 & 28.13 & 28.36 \\
53 & $9.5^{+0.5}_{-3.0}$ & $-0.25^{+0.00}_{-0.00}$ & $7.13^{+0.01}_{-0.10}$ & 3.37 & 28.32 & 27.78 & 27.35 & 27.12 & 26.94 & 27.45 & 27.98 & 28.29 \\
54 & $10.0^{+0.0}_{-0.0}$ & $0.15^{+0.11}_{-0.00}$ & $7.44^{+0.02}_{-0.01}$ & 3.97 & 27.93 & 27.30 & 26.87 & 26.43 & 26.23 & 26.49 & 27.08 & 27.39 \\
55 & $4.0^{+6.0}_{-1.0}$ & $-0.35^{+0.10}_{-0.31}$ & $6.98^{+0.22}_{-0.11}$ & 4.53 & 27.97 & 27.43 & 27.13 & 26.90 & 26.74 & 27.15 & 27.72 & 27.95 \\
56 & $10.0^{+0.0}_{-3.0}$ & $0.15^{+0.00}_{-0.09}$ & $7.43^{+0.01}_{-0.10}$ & 11.05 & 27.81 & 27.31 & 26.81 & 26.47 & 26.31 & 26.71 & 27.21 & 27.43 \\
57 & $7.0^{+2.5}_{-1.0}$ & $0.06^{+0.00}_{-0.31}$ & $7.79^{+0.05}_{-0.06}$ & 4.82 & 26.62 & 26.13 & 25.68 & 25.36 & 25.21 & 25.66 & 26.18 & 26.25 \\
58 & $10.0^{+0.0}_{-3.0}$ & $0.15^{+0.00}_{-0.00}$ & $7.13^{+0.01}_{-0.12}$ & 3.55 & 28.57 & 28.14 & 27.56 & 27.27 & 27.09 & 27.57 & 28.08 & 28.63 \\
59 & $3.8^{+1.2}_{-1.2}$ & $-0.35^{+0.10}_{-0.00}$ & $6.77^{+0.09}_{-0.11}$ & 1.74 & 28.37 & 27.90 & 27.57 & 27.37 & 27.21 & 27.51 & 28.07 & 28.51 \\
60 & $7.0^{+2.5}_{-1.5}$ & $0.06^{+0.00}_{-0.00}$ & $7.06^{+0.06}_{-0.07}$ & 4.97 & 28.60 & 27.89 & 27.49 & 27.22 & 27.03 & 27.06 & 27.35 & 27.94 \\
61 & $10.0^{+0.0}_{-7.5}$ & $-0.66^{+0.31}_{-0.00}$ & $7.03^{+0.01}_{-0.38}$ & 4.20 & 28.30 & 27.87 & 27.55 & 27.37 & 27.23 & 27.54 & 28.17 & 28.54 \\
62 & $8.0^{+1.5}_{-4.0}$ & $-0.66^{+0.00}_{-0.30}$ & $7.03^{+0.05}_{-0.20}$ & 4.57 & 28.01 & 27.66 & 27.35 & 27.18 & 27.09 & 27.58 & 27.97 & 28.15 \\
63 & $9.5^{+0.0}_{-0.0}$ & $0.26^{+0.00}_{-0.00}$ & $6.83^{+0.01}_{-0.01}$ & 9.20 & 29.56 & 28.81 & 28.49 & 28.10 & 27.86 & 27.66 & 27.94 & 28.39 \\
64 & $9.0^{+0.0}_{-0.5}$ & $-1.26^{+0.00}_{-0.23}$ & $7.08^{+0.01}_{-0.02}$ & 4.31 & 27.92 & 27.53 & 27.31 & 27.13 & 27.10 & 27.56 & 28.15 & 28.12 \\
65 & $7.0^{+2.0}_{-1.5}$ & $0.06^{+0.00}_{-0.00}$ & $7.37^{+0.04}_{-0.05}$ & 5.96 & 27.64 & 27.12 & 26.73 & 26.37 & 26.19 & 26.62 & 27.20 & 27.36 \\
66 & $2.5^{+5.5}_{-0.0}$ & $-0.35^{+0.00}_{-0.31}$ & $6.71^{+0.31}_{-0.01}$ & 4.68 & 28.14 & 27.72 & 27.42 & 27.18 & 27.05 & 27.52 & 28.14 & 28.32 \\
67 & $9.0^{+0.5}_{-0.5}$ & $-1.26^{+0.30}_{-0.00}$ & $7.08^{+0.01}_{-0.02}$ & 2.89 & 27.94 & 27.56 & 27.29 & 27.14 & 27.09 & 27.35 & 28.03 & 28.36 \\
68 & $8.5^{+0.0}_{-0.0}$ & $-1.79^{+0.00}_{-0.00}$ & $7.15^{+0.01}_{-0.00}$ & 18.40 & 27.79 & 27.17 & 26.91 & 26.79 & 27.00 & 27.51 & 27.96 & 28.18 \\
69 & $1.0^{+0.0}_{-0.0}$ & $0.26^{+0.00}_{-0.00}$ & $6.78^{+0.01}_{-0.01}$ & 17.90 & 27.39 & 27.39 & 27.16 & 26.97 & 26.19 & 26.50 & 27.44 & 27.18 \\
\enddata
\tablecomments{Observed AB magnitudes are listed in the eight JWST/NIRCam filters used in this study. Ages, metallicities, and stellar masses are listed with uncertainties from Monte Carlo photometric error propagation, where the upper and lower errors correspond to the 84th and 16th percentiles, respectively. The quantity $\chi^2_\nu$ is the reduced chi-square of the best-fit model.}
\end{deluxetable*}

%% file: main.bib
@ARTICLE{H&RCabell2744_2,
       author = {{Harris}, William E. and {Reina-Campos}, Marta},
        title = "{JWST Photometry of Globular Clusters in A2744. II. Luminosity and Color Distributions}",
      journal = {\apj},
         year = 2024,
        month = aug,
       volume = {971},
       number = {2},
          eid = {155},
        pages = {155},
          doi = {10.3847/1538-4357/ad583c},
archivePrefix = {arXiv},
       eprint = {2404.10813},
 primaryClass = {astro-ph.GA},
       adsurl = {https://ui.adsabs.harvard.edu/abs/2024ApJ...971..155H}
}

@ARTICLE{H&RCabell2744_1,
       author = {{Harris}, William E. and {Reina-Campos}, Marta},
        title = "{JWST photometry of globular cluster populations in Abell 2744 at z = 0.3}",
      journal = {\mnras},
         year = 2023,
        month = dec,
       volume = {526},
       number = {2},
        pages = {2696-2708},
          doi = {10.1093/mnras/stad2903},
archivePrefix = {arXiv},
       eprint = {2307.14412},
 primaryClass = {astro-ph.GA},
       adsurl = {https://ui.adsabs.harvard.edu/abs/2023MNRAS.526.2696H}
}

@ARTICLE{UNCOVER,
       author = {{Bezanson}, Rachel and {Labbe}, Ivo and {Whitaker}, Katherine E. and {Leja}, Joel and {Price}, Sedona H. and {Franx}, Marijn and {Brammer}, Gabriel and {Marchesini}, Danilo and {Zitrin}, Adi and {Wang}, Bingjie and {Weaver}, John R. and {Furtak}, Lukas J. and {Atek}, Hakim and {Coe}, Dan and {Cutler}, Sam E. and {Dayal}, Pratika and {van Dokkum}, Pieter and {Feldmann}, Robert and {F{\"o}rster Schreiber}, Natascha M. and {Fujimoto}, Seiji and {Geha}, Marla and {Glazebrook}, Karl and {de Graaff}, Anna and {Greene}, Jenny E. and {Juneau}, St{\'e}phanie and {Kassin}, Susan and {Kriek}, Mariska and {Khullar}, Gourav and {Maseda}, Michael and {Mowla}, Lamiya A. and {Muzzin}, Adam and {Nanayakkara}, Themiya and {Nelson}, Erica J. and {Oesch}, Pascal A. and {Pacifici}, Camilla and {Pan}, Richard and {Papovich}, Casey and {Setton}, David J. and {Shapley}, Alice E. and {Smit}, Renske and {Stefanon}, Mauro and {Taylor}, Edward N. and {Williams}, Christina C.},
        title = "{The JWST UNCOVER Treasury Survey: Ultradeep NIRSpec and NIRCam Observations before the Epoch of Reionization}",
      journal = {\apj},
         year = 2024,
        month = oct,
       volume = {974},
       number = {1},
          eid = {92},
        pages = {92},
          doi = {10.3847/1538-4357/ad66cf},
archivePrefix = {arXiv},
       eprint = {2212.04026},
 primaryClass = {astro-ph.GA},
       adsurl = {https://ui.adsabs.harvard.edu/abs/2024ApJ...974...92B}
}

@ARTICLE{emiles_code,
       author = {{Vazdekis}, A. and {Koleva}, M. and {Ricciardelli}, E. and {R{\"o}ck}, B. and {Falc{\'o}n-Barroso}, J.},
        title = "{UV-extended E-MILES stellar population models: young components in massive early-type galaxies}",
      journal = {\mnras},
         year = 2016,
        month = dec,
       volume = {463},
       number = {4},
        pages = {3409-3436},
          doi = {10.1093/mnras/stw2231},
archivePrefix = {arXiv},
       eprint = {1612.01187},
 primaryClass = {astro-ph.GA},
       adsurl = {https://ui.adsabs.harvard.edu/abs/2016MNRAS.463.3409V}
}

@ARTICLE{BaSTI,
       author = {{Pietrinferni}, Adriano and {Cassisi}, Santi and {Salaris}, Maurizio and {Castelli}, Fiorella},
        title = "{A Large Stellar Evolution Database for Population Synthesis Studies. I. Scaled Solar Models and Isochrones}",
      journal = {\apj},
         year = 2004,
        month = sep,
       volume = {612},
       number = {1},
        pages = {168-190},
          doi = {10.1086/422498},
archivePrefix = {arXiv},
       eprint = {astro-ph/0405193},
 primaryClass = {astro-ph},
       adsurl = {https://ui.adsabs.harvard.edu/abs/2004ApJ...612..168P}
}

@ARTICLE{Ch_IMF,
       author = {{Chabrier}, Gilles},
        title = "{Galactic Stellar and Substellar Initial Mass Function}",
      journal = {\pasp},
         year = 2003,
        month = jul,
       volume = {115},
       number = {809},
        pages = {763-795},
          doi = {10.1086/376392},
archivePrefix = {arXiv},
       eprint = {astro-ph/0304382},
 primaryClass = {astro-ph},
       adsurl = {https://ui.adsabs.harvard.edu/abs/2003PASP..115..763C}
}

@ARTICLE{Isophotal,
       author = {{Tokunaga}, A.~T. and {Vacca}, W.~D.},
        title = "{The Mauna Kea Observatories Near-Infrared Filter Set. III. Isophotal Wavelengths and Absolute Calibration}",
      journal = {\pasp},
         year = 2005,
        month = apr,
       volume = {117},
       number = {830},
        pages = {421-426},
          doi = {10.1086/429382},
archivePrefix = {arXiv},
       eprint = {astro-ph/0502120},
 primaryClass = {astro-ph},
       adsurl = {https://ui.adsabs.harvard.edu/abs/2005PASP..117..421T}
}

@ARTICLE{merger1,
       author = {{Owers}, Matt S. and {Randall}, Scott W. and {Nulsen}, Paul E.~J. and {Couch}, Warrick J. and {David}, Laurence P. and {Kempner}, Joshua C.},
        title = "{The Dissection of Abell 2744: A Rich Cluster Growing Through Major and Minor Mergers}",
      journal = {\apj},
         year = 2011,
        month = feb,
       volume = {728},
       number = {1},
          eid = {27},
        pages = {27},
          doi = {10.1088/0004-637X/728/1/27},
archivePrefix = {arXiv},
       eprint = {1012.1315},
 primaryClass = {astro-ph.CO},
       adsurl = {https://ui.adsabs.harvard.edu/abs/2011ApJ...728...27O}
}

@ARTICLE{merger/substructure1,
       author = {{Jauzac}, M. and {Eckert}, D. and {Schwinn}, J. and {Harvey}, D. and {Baugh}, C.~M. and {Robertson}, A. and {Bose}, S. and {Massey}, R. and {Owers}, M. and {Ebeling}, H. and {Shan}, H.~Y. and {Jullo}, E. and {Kneib}, J. -P. and {Richard}, J. and {Atek}, H. and {Cl{\'e}ment}, B. and {Egami}, E. and {Israel}, H. and {Knowles}, K. and {Limousin}, M. and {Natarajan}, P. and {Rexroth}, M. and {Taylor}, P. and {Tchernin}, C.},
        title = "{The extraordinary amount of substructure in the Hubble Frontier Fields cluster Abell 2744}",
      journal = {\mnras},
         year = 2016,
        month = dec,
       volume = {463},
       number = {4},
        pages = {3876-3893},
          doi = {10.1093/mnras/stw2251},
archivePrefix = {arXiv},
       eprint = {1606.04527},
 primaryClass = {astro-ph.CO},
       adsurl = {https://ui.adsabs.harvard.edu/abs/2016MNRAS.463.3876J}
}

@ARTICLE{substructure2,
       author = {{Eckert}, Dominique and {Jauzac}, Mathilde and {Shan}, Huanyuan and {Kneib}, Jean-Paul and {Erben}, Thomas and {Israel}, Holger and {Jullo}, Eric and {Klein}, Matthias and {Massey}, Richard and {Richard}, Johan and {Tchernin}, C{\'e}line},
        title = "{Warm-hot baryons comprise 5-10 per cent of filaments in the cosmic web}",
      journal = {\nat},
         year = 2015,
        month = dec,
       volume = {528},
       number = {7580},
        pages = {105-107},
          doi = {10.1038/nature16058},
archivePrefix = {arXiv},
       eprint = {1512.00454},
 primaryClass = {astro-ph.CO},
       adsurl = {https://ui.adsabs.harvard.edu/abs/2015Natur.528..105E}
}

@ARTICLE{merger2,
       author = {{Medezinski}, Elinor and {Umetsu}, Keiichi and {Okabe}, Nobuhiro and {Nonino}, Mario and {Molnar}, Sandor and {Massey}, Richard and {Dupke}, Renato and {Merten}, Julian},
        title = "{Frontier Fields: Subaru Weak-Lensing Analysis of the Merging Galaxy Cluster A2744}",
      journal = {\apj},
         year = 2016,
        month = jan,
       volume = {817},
       number = {1},
          eid = {24},
        pages = {24},
          doi = {10.3847/0004-637X/817/1/24},
archivePrefix = {arXiv},
       eprint = {1507.03992},
 primaryClass = {astro-ph.CO},
       adsurl = {https://ui.adsabs.harvard.edu/abs/2016ApJ...817...24M}
}

@ARTICLE{galaxy_pop,
       author = {{Weaver}, John R. and {Cutler}, Sam E. and {Pan}, Richard and {Whitaker}, Katherine E. and {Labb{\'e}}, Ivo and {Price}, Sedona H. and {Bezanson}, Rachel and {Brammer}, Gabriel and {Marchesini}, Danilo and {Leja}, Joel and {Wang}, Bingjie and {Furtak}, Lukas J. and {Zitrin}, Adi and {Atek}, Hakim and {Chemerynska}, Iryna and {Coe}, Dan and {Dayal}, Pratika and {van Dokkum}, Pieter and {Feldmann}, Robert and {F{\"o}rster Schreiber}, Natascha M. and {Franx}, Marijn and {Fujimoto}, Seiji and {Fudamoto}, Yoshinobu and {Glazebrook}, Karl and {de Graaff}, Anna and {Greene}, Jenny E. and {Juneau}, St{\'e}phanie and {Kassin}, Susan and {Kriek}, Mariska and {Khullar}, Gourav and {Maseda}, Michael V. and {Mowla}, Lamiya A. and {Muzzin}, Adam and {Nanayakkara}, Themiya and {Nelson}, Erica J. and {Oesch}, Pascal A. and {Pacifici}, Camilla and {Papovich}, Casey and {Setton}, David J. and {Shapley}, Alice E. and {Shipley}, Heath V. and {Smit}, Renske and {Stefanon}, Mauro and {Taylor}, Edward N. and {Weibel}, Andrea and {Williams}, Christina C.},
        title = "{The UNCOVER Survey: A First-look HST + JWST Catalog of 60,000 Galaxies near A2744 and beyond}",
      journal = {\apjs},
         year = 2024,
        month = jan,
       volume = {270},
       number = {1},
          eid = {7},
        pages = {7},
          doi = {10.3847/1538-4365/ad07e0},
archivePrefix = {arXiv},
       eprint = {2301.02671},
 primaryClass = {astro-ph.GA},
       adsurl = {https://ui.adsabs.harvard.edu/abs/2024ApJS..270....7W}
}

@ARTICLE{Daophot,
       author = {{Stetson}, Peter B.},
        title = "{DAOPHOT: A Computer Program for Crowded-Field Stellar Photometry}",
      journal = {\pasp},
         year = 1987,
        month = mar,
       volume = {99},
        pages = {191},
          doi = {10.1086/131977},
       adsurl = {https://ui.adsabs.harvard.edu/abs/1987PASP...99..191S}
}

@ARTICLE{k_correction,
       author = {{Reina-Campos}, Marta and {Harris}, William E.},
        title = "{RESCUER: cosmological K-corrections for star clusters}",
      journal = {\mnras},
         year = 2024,
        month = jul,
       volume = {531},
       number = {4},
        pages = {4099-4107},
          doi = {10.1093/mnras/stae1414},
archivePrefix = {arXiv},
       eprint = {2310.02307},
 primaryClass = {astro-ph.GA},
       adsurl = {https://ui.adsabs.harvard.edu/abs/2024MNRAS.531.4099R}
}

@ARTICLE{younger_metal_rich,
       author = {{Reina-Campos}, Marta and {Kruijssen}, J.~M. Diederik and {Pfeffer}, Joel L. and {Bastian}, Nate and {Crain}, Robert A.},
        title = "{Formation histories of stars, clusters, and globular clusters in the E-MOSAICS simulations}",
      journal = {\mnras},
         year = 2019,
        month = jul,
       volume = {486},
       number = {4},
        pages = {5838-5852},
          doi = {10.1093/mnras/stz1236},
archivePrefix = {arXiv},
       eprint = {1905.02217},
 primaryClass = {astro-ph.GA},
       adsurl = {https://ui.adsabs.harvard.edu/abs/2019MNRAS.486.5838R}
}

@ARTICLE{lower_ML_ratio,
       author = {{Bell}, Eric F. and {McIntosh}, Daniel H. and {Katz}, Neal and {Weinberg}, Martin D.},
        title = "{The Optical and Near-Infrared Properties of Galaxies. I. Luminosity and Stellar Mass Functions}",
      journal = {\apjs},
         year = 2003,
        month = dec,
       volume = {149},
       number = {2},
        pages = {289-312},
          doi = {10.1086/378847},
archivePrefix = {arXiv},
       eprint = {astro-ph/0302543},
 primaryClass = {astro-ph},
       adsurl = {https://ui.adsabs.harvard.edu/abs/2003ApJS..149..289B}
}

@ARTICLE{two_bands1,
       author = {{Brodie}, Jean P. and {Strader}, Jay},
        title = "{Extragalactic Globular Clusters and Galaxy Formation}",
      journal = {\araa},
         year = 2006,
        month = sep,
       volume = {44},
       number = {1},
        pages = {193-267},
          doi = {10.1146/annurev.astro.44.051905.092441},
archivePrefix = {arXiv},
       eprint = {astro-ph/0602601},
 primaryClass = {astro-ph},
       adsurl = {https://ui.adsabs.harvard.edu/abs/2006ARA&A..44..193B}
}

@ARTICLE{two_bands2,
       author = {{Peng}, Eric W. and {Jord{\'a}n}, Andr{\'e}s and {C{\^o}t{\'e}}, Patrick and {Blakeslee}, John P. and {Ferrarese}, Laura and {Mei}, Simona and {West}, Michael J. and {Merritt}, David and {Milosavljevi{\'c}}, Milos and {Tonry}, John L.},
        title = "{The ACS Virgo Cluster Survey. IX. The Color Distributions of Globular Cluster Systems in Early-Type Galaxies}",
      journal = {\apj},
         year = 2006,
        month = mar,
       volume = {639},
       number = {1},
        pages = {95-119},
          doi = {10.1086/498210},
archivePrefix = {arXiv},
       eprint = {astro-ph/0509654},
 primaryClass = {astro-ph},
       adsurl = {https://ui.adsabs.harvard.edu/abs/2006ApJ...639...95P}
}

@ARTICLE{mass_limit,
       author = {{Harris}, William E. and {Morningstar}, Warren and {Gnedin}, Oleg Y. and {O'Halloran}, Heather and {Blakeslee}, John P. and {Whitmore}, Bradley C. and {C{\^o}t{\'e}}, Patrick and {Geisler}, Douglas and {Peng}, Eric W. and {Bailin}, Jeremy and {Rothberg}, Barry and {Cockcroft}, Robert and {Barber DeGraaff}, Regina},
        title = "{Globular Cluster Systems in Brightest Cluster Galaxies: A Near-universal Luminosity Function?}",
      journal = {\apj},
         year = 2014,
        month = dec,
       volume = {797},
       number = {2},
          eid = {128},
        pages = {128},
          doi = {10.1088/0004-637X/797/2/128},
archivePrefix = {arXiv},
       eprint = {1410.6291},
 primaryClass = {astro-ph.GA},
       adsurl = {https://ui.adsabs.harvard.edu/abs/2014ApJ...797..128H}
}

@ARTICLE{V_band_ratio,
       author = {{McLaughlin}, Dean E. and {van der Marel}, Roeland P.},
        title = "{Resolved Massive Star Clusters in the Milky Way and Its Satellites: Brightness Profiles and a Catalog of Fundamental Parameters}",
      journal = {\apjs},
         year = 2005,
        month = dec,
       volume = {161},
       number = {2},
        pages = {304-360},
          doi = {10.1086/497429},
archivePrefix = {arXiv},
       eprint = {astro-ph/0605132},
 primaryClass = {astro-ph},
       adsurl = {https://ui.adsabs.harvard.edu/abs/2005ApJS..161..304M}
}

@ARTICLE{Harris_two_filters,
       author = {{Harris}, William E.},
        title = "{A Photometric Survey of Globular Cluster Systems in Brightest Cluster Galaxies}",
      journal = {\apjs},
         year = 2023,
        month = mar,
       volume = {265},
       number = {1},
          eid = {9},
        pages = {9},
          doi = {10.3847/1538-4365/acab5c},
archivePrefix = {arXiv},
       eprint = {2212.06174},
 primaryClass = {astro-ph.GA},
       adsurl = {https://ui.adsabs.harvard.edu/abs/2023ApJS..265....9H}
}

@ARTICLE{Previous_SED_fit1,
       author = {{Faisst}, Andreas L. and {Chary}, Ranga Ram and {Brammer}, Gabriel and {Toft}, Sune},
        title = "{What Are Those Tiny Things? A First Study of Compact Star Clusters in the SMACS0723 Field with JWST}",
      journal = {\apjl},
         year = 2022,
        month = dec,
       volume = {941},
       number = {1},
          eid = {L11},
        pages = {L11},
          doi = {10.3847/2041-8213/aca1bf},
archivePrefix = {arXiv},
       eprint = {2208.05502},
 primaryClass = {astro-ph.GA},
       adsurl = {https://ui.adsabs.harvard.edu/abs/2022ApJ...941L..11F}
}

@ARTICLE{Previous_SED_fit2,
       author = {{de Grijs}, R. and {Fritze-v. Alvensleben}, U. and {Anders}, P. and {Gallagher}, J.~S. and {Bastian}, N. and {Taylor}, V.~A. and {Windhorst}, R.~A.},
        title = "{Star cluster formation and evolution in nearby starburst galaxies - I. Systematic uncertainties}",
      journal = {\mnras},
         year = 2003,
        month = jun,
       volume = {342},
       number = {1},
        pages = {259-273},
          doi = {10.1046/j.1365-8711.2003.06536.x},
archivePrefix = {arXiv},
       eprint = {astro-ph/0302286},
 primaryClass = {astro-ph},
       adsurl = {https://ui.adsabs.harvard.edu/abs/2003MNRAS.342..259D}
}

@ARTICLE{GC_MH1,
       author = {{Harris}, William E.},
        title = "{A Catalog of Parameters for Globular Clusters in the Milky Way}",
      journal = {\aj},
         year = 1996,
        month = oct,
       volume = {112},
        pages = {1487},
          doi = {10.1086/118116},
       adsurl = {https://ui.adsabs.harvard.edu/abs/1996AJ....112.1487H}
}

@ARTICLE{GC_MH_age1,
       author = {{Salaris}, M. and {Weiss}, A.},
        title = "{Homogeneous age dating of 55 Galactic globular clusters. Clues to the Galaxy formation mechanisms}",
      journal = {\aap},
         year = 2002,
        month = jun,
       volume = {388},
        pages = {492-503},
          doi = {10.1051/0004-6361:20020554},
archivePrefix = {arXiv},
       eprint = {astro-ph/0204410},
 primaryClass = {astro-ph},
       adsurl = {https://ui.adsabs.harvard.edu/abs/2002A&A...388..492S}
}

@ARTICLE{GC_age1,
       author = {{De Angeli}, Francesca and {Piotto}, Giampaolo and {Cassisi}, Santi and {Busso}, Giorgia and {Recio-Blanco}, Alejandra and {Salaris}, Maurizio and {Aparicio}, Antonio and {Rosenberg}, Alfred},
        title = "{Galactic Globular Cluster Relative Ages}",
      journal = {\aj},
         year = 2005,
        month = jul,
       volume = {130},
       number = {1},
        pages = {116-125},
          doi = {10.1086/430723},
archivePrefix = {arXiv},
       eprint = {astro-ph/0503594},
 primaryClass = {astro-ph},
       adsurl = {https://ui.adsabs.harvard.edu/abs/2005AJ....130..116D}
}

@ARTICLE{GC_M31,
       author = {{Caldwell}, Nelson and {Schiavon}, Ricardo and {Morrison}, Heather and {Rose}, James A. and {Harding}, Paul},
        title = "{Star Clusters in M31. II. Old Cluster Metallicities and Ages from Hectospec Data}",
      journal = {\aj},
         year = 2011,
        month = feb,
       volume = {141},
       number = {2},
          eid = {61},
        pages = {61},
          doi = {10.1088/0004-6256/141/2/61},
archivePrefix = {arXiv},
       eprint = {1101.3779},
 primaryClass = {astro-ph.GA},
       adsurl = {https://ui.adsabs.harvard.edu/abs/2011AJ....141...61C}
}

@ARTICLE{hartman+2025,
       author = {{Hartman}, Kate and {Harris}, William E. and {Kim}, Jinoo},
        title = "{Spectral Energy Distribution Fitting of Globular Clusters in NGC 4874: Masses and Metallicities}",
      journal = {\apj},
         year = 2025,
        month = sep,
       volume = {990},
       number = {1},
          eid = {65},
        pages = {65},
          doi = {10.3847/1538-4357/adf7ac},
archivePrefix = {arXiv},
       eprint = {2508.03684},
 primaryClass = {astro-ph.GA},
       adsurl = {https://ui.adsabs.harvard.edu/abs/2025ApJ...990...65H}
}

@ARTICLE{ML_v_ratio2,
       author = {{Baumgardt}, H. and {Hilker}, M.},
        title = "{A catalogue of masses, structural parameters, and velocity dispersion profiles of 112 Milky Way globular clusters}",
      journal = {\mnras},
         year = 2018,
        month = aug,
       volume = {478},
       number = {2},
        pages = {1520-1557},
          doi = {10.1093/mnras/sty1057},
archivePrefix = {arXiv},
       eprint = {1804.08359},
 primaryClass = {astro-ph.GA},
       adsurl = {https://ui.adsabs.harvard.edu/abs/2018MNRAS.478.1520B}
}

@ARTICLE{ML_v_ratio1,
       author = {{Kimmig}, Brian and {Seth}, Anil and {Ivans}, Inese I. and {Strader}, Jay and {Caldwell}, Nelson and {Anderton}, Tim and {Gregersen}, Dylan},
        title = "{Measuring Consistent Masses for 25 Milky Way Globular Clusters}",
      journal = {\aj},
         year = 2015,
        month = feb,
       volume = {149},
       number = {2},
          eid = {53},
        pages = {53},
          doi = {10.1088/0004-6256/149/2/53},
archivePrefix = {arXiv},
       eprint = {1411.1763},
 primaryClass = {astro-ph.GA},
       adsurl = {https://ui.adsabs.harvard.edu/abs/2015AJ....149...53K}
}

@ARTICLE{GC_UCD_2,
       author = {{Norris}, Mark A. and {Kannappan}, Sheila J. and {Forbes}, Duncan A. and {Romanowsky}, Aaron J. and {Brodie}, Jean P. and {Faifer}, Favio Ra{\'u}l and {Huxor}, Avon and {Maraston}, Claudia and {Moffett}, Amanda J. and {Penny}, Samantha J. and {Pota}, Vincenzo and {Smith-Castelli}, Anal{\'\i}a and {Strader}, Jay and {Bradley}, David and {Eckert}, Kathleen D. and {Fohring}, Dora and {McBride}, JoEllen and {Stark}, David V. and {Vaduvescu}, Ovidiu},
        title = "{The AIMSS Project - I. Bridging the star cluster-galaxy divide$^{★}${\textdagger}{\textdaggerdbl}{\textsection}{\textparagraph}}",
      journal = {\mnras},
         year = 2014,
        month = sep,
       volume = {443},
       number = {2},
        pages = {1151-1172},
          doi = {10.1093/mnras/stu1186},
archivePrefix = {arXiv},
       eprint = {1406.6065},
 primaryClass = {astro-ph.GA},
       adsurl = {https://ui.adsabs.harvard.edu/abs/2014MNRAS.443.1151N}
}

@ARTICLE{GC_UCD_1,
       author = {{Ha{\c{s}}egan}, Monica and {Jord{\'a}n}, Andr{\'e}s and {C{\^o}t{\'e}}, Patrick and {Djorgovski}, S.~G. and {McLaughlin}, Dean E. and {Blakeslee}, John P. and {Mei}, Simona and {West}, Michael J. and {Peng}, Eric W. and {Ferrarese}, Laura and {Milosavljevi{\'c}}, Milo{\v{s}} and {Tonry}, John L. and {Merritt}, David},
        title = "{The ACS Virgo Cluster Survey. VII. Resolving the Connection between Globular Clusters and Ultracompact Dwarf Galaxies}",
      journal = {\apj},
         year = 2005,
        month = jul,
       volume = {627},
       number = {1},
        pages = {203-223},
          doi = {10.1086/430342},
archivePrefix = {arXiv},
       eprint = {astro-ph/0503566},
 primaryClass = {astro-ph},
       adsurl = {https://ui.adsabs.harvard.edu/abs/2005ApJ...627..203H}
}

@ARTICLE{age_mh_degeneracy1,
       author = {{Worthey}, Guy},
        title = "{Comprehensive Stellar Population Models and the Disentanglement of Age and Metallicity Effects}",
      journal = {\apjs},
         year = 1994,
        month = nov,
       volume = {95},
        pages = {107},
          doi = {10.1086/192096},
       adsurl = {https://ui.adsabs.harvard.edu/abs/1994ApJS...95..107W}
}

@ARTICLE{opt_to_NIR,
       author = {{Kissler-Patig}, M. and {Brodie}, J.~P. and {Minniti}, D.},
        title = "{Extragalactic globular clusters in the near infrared. I. A comparison between M87 and NGC 4478}",
      journal = {\aap},
         year = 2002,
        month = aug,
       volume = {391},
        pages = {441-452},
          doi = {10.1051/0004-6361:20020831},
archivePrefix = {arXiv},
       eprint = {astro-ph/0206140},
 primaryClass = {astro-ph},
       adsurl = {https://ui.adsabs.harvard.edu/abs/2002A&A...391..441K}
}

@ARTICLE{vandenberg+2013,
       author = {{VandenBerg}, Don A. and {Brogaard}, K. and {Leaman}, R. and {Casagrande}, L.},
        title = "{The Ages of 55 Globular Clusters as Determined Using an Improved \textbackslashDelta V\^HB\_TO Method along with Color-Magnitude Diagram Constraints, and Their Implications for Broader Issues}",
      journal = {\apj},
         year = 2013,
        month = oct,
       volume = {775},
       number = {2},
          eid = {134},
        pages = {134},
          doi = {10.1088/0004-637X/775/2/134},
archivePrefix = {arXiv},
       eprint = {1308.2257},
 primaryClass = {astro-ph.GA},
       adsurl = {https://ui.adsabs.harvard.edu/abs/2013ApJ...775..134V}
}

@ARTICLE{ying+2025,
       author = {{Ying}, Jiaqi (Martin) and {Chaboyer}, Brian and {Boylan-Kolchin}, Michael and {Weisz}, Daniel R. and {Goebel-Bain}, Rowan},
        title = "{The Absolute Age of Milky Way Globular Clusters}",
      journal = {\apj},
         year = 2025,
        month = jul,
       volume = {987},
       number = {1},
          eid = {52},
        pages = {52},
          doi = {10.3847/1538-4357/add471},
archivePrefix = {arXiv},
       eprint = {2505.02969},
 primaryClass = {astro-ph.GA},
       adsurl = {https://ui.adsabs.harvard.edu/abs/2025ApJ...987...52Y}
}

@ARTICLE{valenzuela+2024,
       author = {{Valenzuela}, Lucas M. and {Remus}, Rhea-Silvia and {McKenzie}, Madeleine and {Forbes}, Duncan A.},
        title = "{Galaxy archaeology for wet mergers: Globular cluster age distributions in the Milky Way and nearby galaxies}",
      journal = {\aap},
         year = 2024,
        month = jul,
       volume = {687},
          eid = {A104},
        pages = {A104},
          doi = {10.1051/0004-6361/202348010},
archivePrefix = {arXiv},
       eprint = {2309.11545},
 primaryClass = {astro-ph.GA},
       adsurl = {https://ui.adsabs.harvard.edu/abs/2024A&A...687A.104V}
}

@ARTICLE{forbes2020,
       author = {{Forbes}, Duncan A.},
        title = "{Reverse engineering the Milky Way}",
      journal = {\mnras},
         year = 2020,
        month = mar,
       volume = {493},
       number = {1},
        pages = {847-854},
          doi = {10.1093/mnras/staa245},
archivePrefix = {arXiv},
       eprint = {2002.01512},
 primaryClass = {astro-ph.GA},
       adsurl = {https://ui.adsabs.harvard.edu/abs/2020MNRAS.493..847F}
}

@ARTICLE{PARSEC,
       author = {{Nguyen}, C.~T. and {Costa}, G. and {Girardi}, L. and {Volpato}, G. and {Bressan}, A. and {Chen}, Y. and {Marigo}, P. and {Fu}, X. and {Goudfrooij}, P.},
        title = "{PARSEC V2.0: Stellar tracks and isochrones of low- and intermediate-mass stars with rotation}",
      journal = {\aap},
         year = 2022,
        month = sep,
       volume = {665},
          eid = {A126},
        pages = {A126},
          doi = {10.1051/0004-6361/202244166},
archivePrefix = {arXiv},
       eprint = {2207.08642},
 primaryClass = {astro-ph.SR},
       adsurl = {https://ui.adsabs.harvard.edu/abs/2022A&A...665A.126N}
}

@ARTICLE{M31_SED,
       author = {{Jiang}, Linhua and {Ma}, Jun and {Zhou}, Xu and {Chen}, Jiansheng and {Wu}, Hong and {Jiang}, Zhaoji},
        title = "{Spectral Energy Distributions and Age Estimates of 172 Globular Clusters in M31}",
      journal = {\aj},
         year = 2003,
        month = feb,
       volume = {125},
       number = {2},
        pages = {727-741},
          doi = {10.1086/345885},
archivePrefix = {arXiv},
       eprint = {astro-ph/0211037},
 primaryClass = {astro-ph},
       adsurl = {https://ui.adsabs.harvard.edu/abs/2003AJ....125..727J}
}

@ARTICLE{M87_SED,
       author = {{Montes}, M. and {Acosta-Pulido}, J.~A. and {Prieto}, M.~A. and {Fern{\'a}ndez-Ontiveros}, J.~A.},
        title = "{The innermost globular clusters of M87}",
      journal = {\mnras},
         year = 2014,
        month = aug,
       volume = {442},
       number = {2},
        pages = {1350-1362},
          doi = {10.1093/mnras/stu948},
archivePrefix = {arXiv},
       eprint = {1405.2920},
 primaryClass = {astro-ph.GA},
       adsurl = {https://ui.adsabs.harvard.edu/abs/2014MNRAS.442.1350M}
}

@ARTICLE{M81_SED,
       author = {{Ma}, Jun and {Zhou}, Xu and {Burstein}, David and {Chen}, Jiansheng and {Jiang}, Zhaoji and {Wu}, Zhenyu and {Wu}, Jianghua},
        title = "{Spectral Energy Distributions of M81 Globular Clusters in the BATC Multicolor Survey}",
      journal = {\pasp},
         year = 2006,
        month = jan,
       volume = {118},
       number = {839},
        pages = {98-106},
          doi = {10.1086/498291},
archivePrefix = {arXiv},
       eprint = {astro-ph/0512173},
 primaryClass = {astro-ph},
       adsurl = {https://ui.adsabs.harvard.edu/abs/2006PASP..118...98M}
}

@ARTICLE{Virgo_SED,
       author = {{Powalka}, Mathieu and {Lan{\c{c}}on}, Ariane and {Puzia}, Thomas H. and {Peng}, Eric W. and {Liu}, Chengze and {Mu{\~n}oz}, Roberto P. and {Blakeslee}, John P. and {C{\^o}t{\'e}}, Patrick and {Ferrarese}, Laura and {Roediger}, Joel and {S{\'a}nchez-Janssen}, R{\'u}ben and {Zhang}, Hongxin and {Durrell}, Patrick R. and {Cuillandre}, Jean-Charles and {Duc}, Pierre-Alain and {Guhathakurta}, Puragra and {Gwyn}, S.~D.~J. and {Hudelot}, Patrick and {Mei}, Simona and {Toloba}, Elisa},
        title = "{The Next Generation Virgo Cluster Survey (NGVS). XXV. Fiducial Panchromatic Colors of Virgo Core Globular Clusters and Their Comparison to Model Predictions}",
      journal = {\apjs},
         year = 2016,
        month = nov,
       volume = {227},
       number = {1},
          eid = {12},
        pages = {12},
          doi = {10.3847/0067-0049/227/1/12},
archivePrefix = {arXiv},
       eprint = {1608.04742},
 primaryClass = {astro-ph.GA},
       adsurl = {https://ui.adsabs.harvard.edu/abs/2016ApJS..227...12P}
}

@ARTICLE{Padova00,
       author = {{Girardi}, L. and {Bressan}, A. and {Bertelli}, G. and {Chiosi}, C.},
        title = "{Evolutionary tracks and isochrones for low- and intermediate-mass stars: From 0.15 to 7 M$_{sun}$, and from Z=0.0004 to 0.03}",
      journal = {\aaps},
         year = 2000,
        month = feb,
       volume = {141},
        pages = {371-383},
          doi = {10.1051/aas:2000126},
archivePrefix = {arXiv},
       eprint = {astro-ph/9910164},
 primaryClass = {astro-ph},
       adsurl = {https://ui.adsabs.harvard.edu/abs/2000A&AS..141..371G}
}

@ARTICLE{high_chi_square_explanation,
       author = {{Proctor}, Robert N. and {Forbes}, Duncan A. and {Beasley}, Michael A.},
        title = "{A robust method for the analysis of integrated spectra from globular clusters using Lick indices}",
      journal = {\mnras},
         year = 2004,
        month = dec,
       volume = {355},
       number = {4},
        pages = {1327-1338},
          doi = {10.1111/j.1365-2966.2004.08415.x},
archivePrefix = {arXiv},
       eprint = {astro-ph/0409526},
 primaryClass = {astro-ph},
       adsurl = {https://ui.adsabs.harvard.edu/abs/2004MNRAS.355.1327P}
}

@ARTICLE{high_chi_square_value,
       author = {{Ma}, Jun and {Yang}, Yanbin and {Burstein}, David and {Fan}, Zhou and {Wu}, Zhenyu and {Zhou}, Xu and {Wu}, Jianghua and {Jiang}, Zhaoji and {Chen}, Jiansheng},
        title = "{Age Constraints for an M31 Globular Cluster from SEDs Fit}",
      journal = {\apj},
         year = 2007,
        month = apr,
       volume = {659},
       number = {1},
        pages = {359-364},
          doi = {10.1086/511850},
archivePrefix = {arXiv},
       eprint = {astro-ph/0612368},
 primaryClass = {astro-ph},
       adsurl = {https://ui.adsabs.harvard.edu/abs/2007ApJ...659..359M}
}

@ARTICLE{high_chi_square_adopt,
       author = {{Fan}, Zhou and {de Grijs}, Richard and {Chen}, Bingqiu and {Jiang}, Linhua and {Bian}, Fuyan and {Li}, Zhongmu},
        title = "{Lick Indices and Spectral Energy Distribution Analysis Based on an M31 Star Cluster Sample: Comparisons of Methods and Models}",
      journal = {\aj},
         year = 2016,
        month = dec,
       volume = {152},
       number = {6},
          eid = {208},
        pages = {208},
          doi = {10.3847/0004-6256/152/6/208},
archivePrefix = {arXiv},
       eprint = {1610.07002},
 primaryClass = {astro-ph.GA},
       adsurl = {https://ui.adsabs.harvard.edu/abs/2016AJ....152..208F}
}
